\documentclass[aps,preprint,superscriptaddress,preprintnumbers,floatfix]{revtex4-1}

\usepackage{amsthm}
\usepackage{amsmath}
\usepackage{mathrsfs}
\usepackage{amssymb}
\usepackage{wasysym}
\usepackage{graphicx}
\usepackage{varwidth}
\usepackage{url}
\usepackage{dsfont}
\usepackage{bm}
\usepackage{ifthen}
\usepackage[usenames,dvipsnames]{color}
\usepackage{mathrsfs}

\usepackage{physics}
\usepackage{xcolor}
\usepackage{fancyhdr}
\usepackage[normalem]{ulem}
\usepackage{mathtools}
\usepackage[margin=1in]{geometry}
\usepackage{xcolor}
\usepackage{tikz}
\usetikzlibrary{arrows.meta,decorations.markings}
\usepackage[colorlinks=true,linkcolor=blue!55!black,citecolor=blue!55!black]{hyperref}


\usepackage{enumitem}
\usepackage{mathtools}
\usepackage{algorithm,algpseudocode}
\algnewcommand{\LeftComment}[1]{\Statex \hspace{3em}\(\triangleright\) #1}
\usepackage{multirow}
\usepackage{booktabs}
\usepackage{siunitx} 
\newcommand{\Ei}{\mathrm{Ei}}

\newcommand{\betaf}{\beta_{\mathrm{f}}}
\newcommand{\Jf}{\mathcal J_{\mathrm{f}}}

\newcommand{\e}{\mathrm{e}}

\usepackage{amsthm}

\theoremstyle{definition}

\begin{document}

\title{Folded out-of-time-order correlators in double-scaled SYK}

\begin{abstract}
Harlow and Zhao~\cite{harlow2026antiscramblingeuclideanfoldsobserver}
have identified ``anti-scrambling'' as a characteristic signature of
cosmological horizons in de Sitter space, and proposed that it can be
realized in a quantum system with bounded energy by Euclidean time
``folding" of out-of-time-order correlators (OTOCs).
We implement this prescription in the double-scaled Sachdev-Ye-Kitaev
(DSSYK) model. We
use~\cite{mukhametzhanov2023largepsykchord} to
define an infinite-temperature ($\beta$$=$$0$) DSSYK OTOC 
that in the semiclassical limit of small values
of the double-scaling parameter $\lambda$ reproduces the usual anti-de Sitter eikonal shockwave behavior with its characteristic time delay, and then apply the Harlow-Zhao fold with the ``fake'' DSSYK inverse temperature $\beta_f$ playing the role of $\beta$. 
We demonstrate both numerically and analytically that the resulting folded OTOC displays anti-scrambling. The exact DSSYK evaluation tracks the semiclassical de Sitter eikonal formula in the scrambling regime, including a time advance instead of a time delay. 
The direction of the fold selects the side from which the continued
eikonal amplitude approaches its branch cut.
We show how our results are consistent with the ``no-go" constraints
of~\cite{cui2026sitterantiscramblingalgebra,chen2026negativeshocksversusstatic}.
Using the face energies of the chord expansion, we compare the folded
correlator with the two-shell analysis
of~\cite{harlow2026antiscramblingeuclideanfoldsobserver}: the growing
term of the two-shell dilaton invariant appears in DSSYK with the same
exponent, but as an imaginary part, while its real part,
which would signal the backreaction that ends in a crunch, only moves by
$O(\lambda)$ for thermal-scale probes. For heavy probes the correlator
factorizes and the folded face is a thermal ensemble of DSSYK at
negative temperature continued in real
time, whose energy reaches the
upper edge of the spectrum, which is the Nariai point of the sine-dilaton dual, at a time set by $\beta_f$.
We also generalize the folding prescription to finite $\beta$; we find
reasonable agreement with the de Sitter eikonal form
for sufficiently high temperatures, but at lower temperatures
the folded correlator is dominated by a
pair of complex conjugate saddles, an anti-Stokes crossover that prevents retrieval of
the eikonal behavior from the folded OTOC.
\end{abstract}

\author{Joseph D. Lykken}
\affiliation{Quantum Division, Fermi National Accelerator Laboratory, Batavia, IL, USA}

\author{Madeline Mooney}
\affiliation{Division of Physics, Mathematics and Astronomy, Caltech, Pasadena, CA, USA}

\author{Maria Spiropulu}
\affiliation{Division of Physics, Mathematics and Astronomy, Caltech, Pasadena, CA, USA}

\maketitle

\section{Introduction}

Our understanding of black hole physics and the holographic AdS/CFT correspondence have been significantly advanced by study of the
Sachdev-Ye-Kitaev (SYK) model~\cite{PhysRevLett.70.3339,
kitaev2015simple,Sachdev:2015efa,Maldacena:2016hyu,garcia2016spectral,
PhysRevD.96.066012,Kitaev2018,Roberts:2018mnp}.
In a certain parameter regime
corresponding to low energy and low temperature, 
the model exhibits an approximate
conformal behavior holographically dual to
Jackiw-Teitelboim (JT) two-dimensional gravity.
Furthermore SYK correlators exhibit thermalization and scrambling
behavior related to the physics of probing the near horizon region
of a black hole~\cite{Maldacena:2001kr,Son_2002,
Shenker:2013pqa,Roberts_2015,shenker2015stringyeffectsscrambling}.

Equally insightful is the double-scaled version of the SYK
model (DSSYK)~\cite{Cotler2017,Berkooz_2018,Berkooz:2018jqr,
Berkooz:2020uly,Lin_2023,mukhametzhanov2023largepsykchord},
which is fully solvable using the chord diagram method,
and whose holographic dual has been identified as
sine-dilaton gravity~\cite{Blommaert_2025}.

It would be helpful to have a similarly tractable and insightful
model to study the cosmological horizon of de Sitter space and make
progress on the notoriously difficult problem of defining and understanding
de Sitter holography
~\cite{witten2001quantumgravitysitterspace,Strominger_2001,Maldacena_2003}.
Susskind and other authors have investigated the
possibility that perhaps DSSYK itself, at infinite temperature,
can be used to some extent as a testbed for de Sitter horizon physics
~\cite{susskind2021entanglementchaossitterholography,
susskind2022scramblingdoublescaledsyksitter,
lin2022infinitetemperatureshot,
rahman2024dsjtgravitydoublescaled,
susskind2022sitterspacedoublescaledsyk,
narovlansky2023doublescaledsyksitterholography,
xu2026vonneumannalgebrasdoublescaled,
rahman2025infinitetemperatureinfinitetemperatures}.
Harlow and Zhao advanced this prospect by proposing a concrete
prescription for ``folding" out-of-time-order (OTOC) 4-point correlators
of a quantum system with bounded energy, so as to obtain OTOCs
relevant to probing de Sitter
cosmological horizons, and suggested testing their
folding prescription in the SYK model.

Here we will implement the folding 
of ~\cite{harlow2026antiscramblingeuclideanfoldsobserver} 
on OTOCs of the
DSSYK model, whose bounded spectrum makes the backward Euclidean
evolution required by the fold well-defined.
We make one key modification of their protocol:
instead of applying it to thermal Euclidean OTOCs at some inverse
temperature $\beta$, we apply it to $\beta=0$ OTOCs with
the DSSYK ``fake" inverse temperature $\beta_f$ defining the Euclidean thermal circle.
The fake temperature is a natural choice given the results 
of 
~\cite{susskind2021entanglementchaossitterholography,
susskind2022scramblingdoublescaledsyksitter,
lin2022infinitetemperatureshot,
rahman2024dsjtgravitydoublescaled,
susskind2022sitterspacedoublescaledsyk,
narovlansky2023doublescaledsyksitterholography}
where the fake temperature is identified with the de Sitter temperature.

Using chord diagrams we demonstrate explicitly that the modified
DSSYK OTOCs exhibit ``anti-scrambling", a characteristic feature
of probing near a de Sitter cosmological horizon, related to a time
advance rather than the time delay characteristic of
probing near a black hole horizon in AdS. The folded OTOCs match
to the expected de Sitter eikonal behavior as one approaches
the scrambling time,
with the Lyapunov exponent of the $\beta=0$ theory, $\lambda_L=2\pi/\betaf$,
which is the value the chaos bound~\cite{maldacena2016bound} would allow
at a temperature $1/\betaf$.
The sheet of the analytically continued eikonal amplitude is fixed by
the direction of the fold as confirmed by the exact DSSYK
evaluation.

The fold itself defines the fake period at finite $\lambda$ 
and the result agrees at NLO with the renormalized period of the two-point function.
We also study the folded OTOC's behavior at later times,
beyond twice the scrambling time, where the analysis
of~\cite{harlow2026antiscramblingeuclideanfoldsobserver} observed 
a late time de Sitter crunch.

In Section II we review the basics of the de Sitter dynamics and the
proposal of~\cite{harlow2026antiscramblingeuclideanfoldsobserver}. 
In Sections III and IV we review
the DSSYK model, its holographic dual, and the idea of fake temperature. In Section V
we describe how to obtain the folded DSSYK OTOCs, and in Sections VI and VII we analyze their properties. Section VIII has a
comparison of the folded correlator with the two-shell
classical analysis
of~\cite{harlow2026antiscramblingeuclideanfoldsobserver}.
In Section IX we explain how our results are consistent with
the ``no-go'' constraints obtained
in~\cite{cui2026sitterantiscramblingalgebra,chen2026negativeshocksversusstatic}, and suggest some topics for further investigation.

\section{de Sitter cosmological horizons and anti-scrambling}

The scattering of ingoing and outgoing particles with a large
relative boost near a black
hole horizon can be described in terms of the eikonal approximation
as a particle interacting with a gravitational 
shock wave~\cite{Dray:1984ha}. Suppose $W$ is an operator
creating a particle at some time $t=0$ moving close to the
horizon with lightcone momentum $q_-$, and let the operator $V$ create
a probe at a later time $t$ with lightcone momentum $p_+$. Looking at
the 1+1 dimensional case, the probe
crossing the boosted gravitational shock wave 
has the effect of abruptly shifting its Kruskal lightcone coordinate by
some amount $x\,$e$^{2\pi t/\beta}$$q_-$, where $x$ is positive and proportional to Newton's constant $G_N$, and e$^{2\pi t/\beta}$
is the near horizon
exponential boost factor, whose rate is set by the Hawking
temperature $1/\beta$. This shift is called the
time delay. Quantum mechanically the effect on the probe's wavefunction
is that it picks up the eikonal phase factor e$^{ia(t)p_+q_-}$
where $a(t)$ = $x\,$e$^{2\pi t/\beta}$; this is known as the 1+1 dimensional eikonal $S$ matrix.

The effect is a time delay because the shock's positive energy causes the black hole horizon size to grow; the near horizon probe finds itself
closer to the horizon and its clock slows down.
We note in passing that a negative energy shock wave, as is engineered for
the Gao-Jafferis-Wall traversable wormhole~\cite{gao2017traversable},
shrinks the horizon and causes a time advance.

Continuing to Euclidean time converts the eikonal $S$ matrix into a
real exponential whose expectation value we could in principle
compute:
\begin{align}
\Phi(a) = \langle {\rm e}^{-a(t)p_+q_-} \rangle
= \int_0^\infty dp \int_0^\infty dq \,\rho_V (p) \rho_W (q)
e^{-a(t)p q}
\end{align}
where in the second equality we introduced explicit
momentum wavefunctions $\rho_V$,$\rho_W$.

Focusing now on JT gravity in the approximately
conformal Schwarzian picture, we
can write $a(t) \sim (\Delta_V\Delta_W/C)\,$e$^{2\pi t/\beta}$,
where $\Delta_V$,$\Delta_W$ are the conformal dimensions of the
probes, and $C$ is the Schwarzian coupling. Taking
$\Delta_V$=$\Delta_W$ =1, the Fourier/Legendre transform
of the conformal 2-point correlators gives:
\begin{align}\label{eq:momd}
\rho_V(p) = p\,{\rm e}^{-p}\;;\quad
\rho_W(q) = q\,{\rm e}^{-q}
\end{align}
from which we get the eikonal function
\begin{align}\label{eq:Phi}
\Phi(a) = \int_0^\infty dp\, \frac{p\,{\rm e}^{-p}}
{(1+ap)^2}
=\frac{1}{a^2}\,U(2,1,\frac{1}{a})
= w^2\Big[(1+w)e^{w}E_1(w)-1\Big],\qquad w=1/a\;
\end{align}
where $U$ is a confluent hypergeometric function, $E_1$ is the
exponential integral, and $\Phi$ is normalized such that
$\Phi(a) \to 1$ as $a \to 0$; its asymptotic expansion is
$\Phi(a) = 1 - 4a + 18a^2 - 96 a^3 + \dots$; $\Phi(a)$ is exactly the function
$F(w)\equiv -4w^2 I(w)$ of~\cite{harlow2026antiscramblingeuclideanfoldsobserver}
(their equation (4.32)), with $w = 1/a$.

As reviewed in \cite{harlow2026antiscramblingeuclideanfoldsobserver} 
when we probe de Sitter cosmological
horizons the time delay discussed above becomes a time advance.
This is because a de Sitter observer views the cosmological horizon
``from the inside", and a positive energy particle moving through a de Sitter horizon moves the horizon away from the observer rather than towards them.

In the eikonal analysis above, Harlow and Zhao as well as
Cui and Kolchmeyer
~\cite{cui2026sitterantiscramblingalgebra}
observe that
one can get the de Sitter version by simply taking
$\Phi(a(t)) \to \Phi(-a(t))$. This implies a time advance
in the shockwave picture rather than a time delay.
The integral in equation (\ref{eq:Phi}) then has a
double pole on the contour at $p=1/a$, and the double integral
over $p$ and $q$ no longer converges: the ``attractive''
$e^{+apq}$ beats the $e^{-(p+q)}$ suppression at large momenta.
$\Phi(-a)$ is instead defined by
continuing the closed form, $w\to-w$, with $E_1(-w)=-\Ei(w)\mp i\pi$;  the two signs correspond to the two sides of the branch cut of $E_1$ along the negative real axis. 
The sign is not fixed by the replacement $\Phi(a)\to\Phi(-a)$ alone;
we will see in Section~\ref{sec:rotation} that each direction of fold
approaches the cut from a definite side, confirmed by
the exact DSSYK numerics of Section~\ref{sec:numerics}.

Thus in the de Sitter case $\Phi$ is complex;
the real part is common to both sheets:
\begin{equation}\label{eq:phiex}
\mathrm{Re}\,\Phi(-a)=w^2\Big[(1-w)e^{-w}\big(-\Ei(w)\big)-1\Big]=1+4a+18a^2+96a^3+\dots,
\end{equation}
all terms positive. The imaginary part is
$\pm\pi w^2(w-1)e^{-w}$, which is exponentially small when
$a$ is small. For sufficiently small $a$ we can also
approximate the de Sitter $\Phi$ from the first terms
in the asymptotic series $1+4a+18a^2+\dots$.

The consequence is that the leading correction at early times
has the opposite sign
from the black-hole case, i.e.\ it initially grows
instead of decaying. Harlow and Zhao call this
anti-scrambling.  
Cui and Kolchmeyer~\cite{cui2026sitterantiscramblingalgebra}
(see also \cite{milekhin2026outoftimeorderedcorrelatorssitterrevisited,
chen2026negativeshocksversusstatic})
showed
that the time advance is incompatible with the KMS condition for the
vacuum correlators of the observer's matter operators, and that,
once the observer's clock is quantized, the
Hartle--Hawking state is not a trace on the observer's algebra.

A way around this was proposed 
in~\cite{harlow2026antiscramblingeuclideanfoldsobserver}. 
In a quantum
system whose Hamiltonian is bounded above and below, backward Euclidean evolution $e^{+\tau H}$
exists, and one can in principle define correlators whose Euclidean contour is ``folded''.  Their prescription: take the
thermal four-point OTOC $\Tr[e^{-\beta H}V(\tau_4)W(\tau_3)V(\tau_2)W(\tau_1)]$, $0<\tau_1<\dots<\tau_4<\beta$, and replace
\begin{equation}
(\tau_1,\tau_2,\tau_3,\tau_4)\;\longrightarrow\;(-\tau_1,\ n\beta-\tau_2,\ \beta-\tau_3,\ (n+1)\beta-\tau_4),\qquad n\in\mathbb Z .
\label{eq:HZ}
\end{equation}
The integer $n$ is
undetermined in the semiclassical analysis; at finite $\lambda$ different $n$ correspond to different correlators; here we consider the two minimal folds $n=0,1$, which turn out to be complex conjugates of each other (Section~\ref{sec:fold}).
They note 
that the prescription assumes the analytic
continuation commutes with the semiclassical limit, that it could fail through a Stokes phenomenon (a point we address in Section~\ref{sec:fail}), 
and that
it covers only the regime where the backreaction
is ``mild". They suggested
testing their prescription in the SYK model.


\section{DSSYK and sine-dilaton gravity}\label{sec:sinedg}
The chord diagram approach to DSSYK is nicely reviewed
by Berkooz and Mamroud~\cite{berkooz2024cordialintroductiondoublescaled}.
The $p$-local SYK Hamiltonian is
\begin{align}
H=i^{p/2}\sum_{I}J_I\,\Psi_I
\end{align}
where the $I$ denote ordered strings of $p$ indices 
${i_1,i_2,\ldots i_p}$ ranging over $N$ possible values,
the $J_I$ denote Gaussian random couplings times an overall
energy scale $J$, and $\Psi_I$ denotes a string of Majorana fermions
with indices in $I$. DSSYK is defined as the
double-scaled limit $N,p\to\infty$ with
$\lambda=2p^2/N$ fixed and $q=e^{-\lambda}$. 
The Gaussian
couplings have variance chosen so that $\Tr[H^2]=J^2$, where
we use the normalized trace $\Tr[\mathds{1}] = 1$.
The disorder-averaged moments are chord-diagram sums,
\begin{equation}
\Tr[H^{2k}]=J^{2k} \mspace{-25mu}
\sum_{\text{pairings of }2k\text{ points}} \mspace{-25mu}
q^{\#\text{crossings}},
\end{equation}
The energy spectrum is given by
\begin{align}\label{eq:energy}
E(s) = -E_0\,{\rm cos}(\lambda s)\,, \qquad s\in [0,\pi/\lambda ]
\,, \qquad
E_0= \frac{2J}{\sqrt{1-q}}\, ,
\end{align}
thus the spectrum is bounded from above and below by $\pm E_0$;
this is what makes DSSYK a
candidate for the construction of~\cite{harlow2026antiscramblingeuclideanfoldsobserver}, 
which requires
backward Euclidean evolution $e^{+\tau H}$ to exist.

The holographic bulk dual of DSSYK has been identified as 
two-dimensional sine-dilaton gravity~\cite{Blommaert_2025}.
With our conventions, and setting $8\pi G_N$$=$$1$, the sine-dilaton action is
given by:
\begin{align}\label{eq:sdaction}
\frac{1}{\lambda}\left[
\frac{1}{2}\int\,d^2x\,\sqrt{g}\,
(\Phi_{\rm sd} R +2\,{\rm sin}\,\Phi_{\rm sd}) + \text{(boundary terms)} \right]
\end{align}
where $\Phi_{\rm sd}$ is the dilaton field of~\cite{Blommaert_2025} after their rescaling, 
and $R$ is the Ricci scalar of the metric $g$. In the Schwarzschild gauge
the classical solutions of sine-dilaton gravity are parameterized by an
angle $\theta$ and can be written:
\begin{align}\label{eq:sdsol}
ds^2 = -F\,dt^2 + \frac{dr^2}{F}\,,
\qquad \Phi_{\rm sd} = r\,,
\qquad
F(r) = 2\,({\rm cos}\,\theta - {\rm cos}\,r)
\end{align}
The ADM energy of these solutions is given by
$E = -E_0\,{\rm cos}\,\theta$, with the boundary time normalized as for
the DSSYK Hamiltonian above.
These solutions have a black hole horizon at $r = \theta$ and
a cosmological horizon at $r = 2\pi - \theta$.

Like other dilaton gravities, sine-dilaton gravity has a conserved
charge, constant throughout a region without matter; it can be
written as:
\begin{align}\label{eq:kappasd}
\kappa_{\rm sd}(\theta) = (\nabla \Phi_{\rm sd})^2 
+ 2\,{\rm cos}\,\Phi_{\rm sd}
= 2\,{\rm cos}\,\theta
\end{align}

DSSYK is a complete quantum
theory of two-dimensional dilaton gravity whose spectrum is bounded above and below. In the low temperature limit with $\lambda \to 0$,
$s$ fixed, DSSYK reduces to the quantum Schwarzian system
dual to JT gravity. 

In a different limit
$\lambda \to 0$, $\lambda s$ fixed, $\beta = x/\sqrt\lambda$ with $x$ fixed, and $J$$=$$1$,
the DSSYK partition
function can be estimated using a 
saddle point~\cite{mukhametzhanov2023largepsykchord,Maldacena:2016hyu}.
The partition function is given by (see
eqs.~(3.8) and (3.9) of~\cite{mukhametzhanov2023largepsykchord}
\footnote{note that Mukhametzhanov normalizes the
Hamiltonian such that $\Tr[H^2] = 1/\lambda$, so that his energies
and times are related to ours by $E\to E/\sqrt\lambda$, $\beta \to \sqrt{\lambda}\,\beta$, $u\to\sqrt\lambda\,u$.}):
\begin{align}\label{eq:Zdef}
Z(\beta ) = \int_0^{\pi/\lambda} ds\,\rho_q(s)\,
\e^{-\beta \, E(s)}
\end{align}
where, to leading order in small $\lambda$,
$-\beta\,E(s) = (2x/\lambda)$cos$(\theta)$, where
$s = \theta/\lambda$, with $\theta \in [0,\pi ]$.
\begin{align}\label{eq:dstates}
\rho_q(s) \propto \e^{\mathcal S(\theta)/\lambda}\; ;\qquad
\mathcal S(\theta) = 2(\pi \theta - \theta^2)
\end{align}
We see that the total exponent
$[\mathcal S(\theta)+2x\,{\rm cos}\,\theta]/\lambda$ is of order
$1/\lambda$, with a saddle point where its derivative with respect to
$\theta$ vanishes, giving:
\begin{align}\label{eq:saddle}
\frac{v}{2\,{\rm cos}\frac{\pi}{2}v}
=\frac{x}{2\pi}\,,\qquad
v = 1 - \frac{2}{\pi}\theta
\end{align}
Expanding the exponent to second order about the saddle $\theta_s$, the thermal distribution of $\theta$ is Gaussian with variance:
\begin{align}\label{eq:variance}
\sigma_\theta^2 = \frac{\lambda}{4+2x\,\mathrm{cos}\,\theta_s}
\end{align}
We will be interested in the infinite temperature case
$\beta = 0$, where the saddle point is at the midpoint $\theta = \pi/2$
of the energy spectrum, and the variance reduces to $\lambda/4$.


\section{Fake temperature in DSSYK}\label{sec:fake}

The fake temperature of DSSYK appears when we consider
two-point functions of ``matter" operators.
We introduce an unnormalized basis of chord states $|n\rangle$, $n\ge0$, labeled by the number of open chords, and the transfer matrix
\begin{equation}
T|n\rangle=|n+1\rangle+[n]_q\,|n-1\rangle,\qquad [n]_q=\frac{1-q^n}{1-q},
\end{equation}
where $|n+1\rangle$ opens a chord (weight 1) and $[n]_q=\sum_{j=0}^{n-1}q^{j}$ closes one of the $n$ open chords,
summed over which one, each crossing the $j$ chords opened after it.  The vacuum-to-vacuum matrix element
$\langle0|T^{2k}\ket{0}$, with $\bra{0}$ the functional extracting the $|0\rangle$ coefficient, is then exactly the
chord-diagram sum, so the Hamiltonian restricted to the chord Hilbert space is $H=J\,T$, with $\bra{0} T^2 \ket{0} = 1$.

Introduce now some matter operator $V$, which is a sum
over strings of $\tilde{p}$ Majoranas:
\begin{align}
V = i^{\tilde{p}/2}\sum_I'\, \bar{J}_{I'}\,\Psi_{I'}
\end{align}
where the $\bar{J}_{I'}$ are Gaussian random couplings.
We now have to keep track of two kinds of chords: H-chords related
to the Hamiltonian, and matter chords related to $V$.
The intersection of two H-chords has weight $q$, but the
intersection of an H-chord with a matter chord has
weight $q^\Delta$, where $\Delta = \tilde{p}/p$ and in
the semiclassical conformal regime corresponds to the conformal dimension
of $V$.
 
The Euclidean two-point function of $V$ is
\begin{equation}
G(\tau)=\frac{\bra{0}e^{-(\beta-\tau)T}q^{\Delta\hat n}e^{-\tau T}\ket{0}}{Z(\beta)}
\end{equation}
where $\hat n$ is the chord number operator
$\hat n \ket{n} = n\ket{n}$, $Z(\beta )$ is the partition function and $\beta$ is the inverse
temperature.
From now on we will be focused on DSSYK correlators in the infinite
temperature limit $\beta\to 0$. 
Switching to Lorentzian time $t$ and letting $G_0(it )$ denote
the $\beta = 0$ correlator and using $Z(0) = 1$, we have:
\begin{equation}\label{eq:G0}
G_0(it)=\bra{0}e^{it T}q^{\Delta\hat n}e^{-it T}\ket{0}
\end{equation}

In the semiclassical limit 
$\lambda\to0$ at fixed $\mathcal Jt$, 
$G_0(it)$ reduces to the form~\cite{Maldacena:2016hyu,berkooz2024cordialintroductiondoublescaled}
\begin{equation}\label{eq:G0sc}
G^{\rm sc}_0(it)=\cosh(\mathcal Jt)^{-2\Delta}
\end{equation}
We can determine $\mathcal{J}$ by expanding (\ref{eq:G0})
to leading order in $t$ and matching to (\ref{eq:G0sc})
expanded to the same order:
\begin{equation}\label{eq:matchJ}
\mathcal J^2=J^2\,\frac{1-q^{\Delta}}{\Delta}=J^2\lambda\Big(1-\frac{\Delta}{2}\lambda+\dots\Big),
\end{equation}
i.e.\ $\mathcal J=\sqrt{\lambda}J$ at leading order.
From now on we will set $J=1$.

We can write $G_0(it)$ as the Fourier transform of a spectral
density $\rho_\Delta$:
\begin{equation}\label{eq:FT}
G_0(it)=\int\frac{d\omega}{2\pi}\,\rho_\Delta(\omega)\,e^{-i\omega t},\qquad
\rho_\Delta(\omega)=\frac{2\pi}{D}\sum_{ij}|V_{ij}|^2\,\delta(\omega-E_i+E_j),
\end{equation}
where $D$ is the dimension of the Hilbert space.
In the semiclassical limit the Fourier transform is a Beta-function integral, giving:
\begin{equation}
\rho^{sc}_\Delta(\omega)=\int dt\,e^{i\omega t}\cosh^{-2\Delta}(\mathcal Jt)
=\frac{2^{2\Delta-1}}{\mathcal J\,\Gamma(2\Delta)}\Big|\Gamma\!\Big(\Delta+\frac{i\omega}{2\mathcal J}\Big)\Big|^2 .
\label{eq:rhosemi}
\end{equation}
Using Stirling's formula $|\Gamma(\Delta+ix)|^2\simeq2\pi|x|^{2\Delta-1}e^{-\pi|x|}$ as $|x|\to\infty$
with $x=\omega/2\mathcal J$, we get an expression for the tails
of the spectral density:
\begin{equation}\label{eq:tails}
\rho_\Delta(\omega)\;\propto\;|\omega|^{2\Delta-1}\,e^{-\pi|\omega|/2\mathcal J}
\end{equation}
In the variable 
$p=\pi \omega/(2{\mathcal J})$ this is $\propto p^{2\Delta-1}e^{-p}$.
Comparing with the eikonal momentum distributions
of (\ref{eq:momd}) we see that they are
related to the tails of the spectral function.

Using (\ref{eq:G0sc}) and (\ref{eq:matchJ}) we can define a fake inverse
temperature $\beta_f = \pi/{\mathcal J} \simeq \pi/\sqrt{\lambda}$
(at finite $\lambda$ the coupling that sets the fake period will differ from
the short-time coupling $\mathcal J$ of (\ref{eq:matchJ}) at $O(\lambda)$; we
return to this below and write $\betaf \equiv \pi/\Jf$ for the former):
\begin{align}\label{eq:betaf}
G^{\rm sc}_0(\tau)&=\cos\left(\frac{\pi\tau}{\beta_f}\right)^{-2\Delta}
= \sin\left[\frac{\pi}{\beta_f} (\tau + \beta_f/2)\right]^{-2\Delta}
\end{align}
Rewriting the tails of the spectral density as
\begin{equation}\label{eq:tailsb}
\rho_\Delta(\omega)\;\propto\;|\omega|^{2\Delta-1}\,e^{-\beta_f|\omega|/2}
\end{equation}
and examining the convergence for complex $t=t_r-i\tau$ of the integrand in (\ref{eq:FT}), we see that convergence at
$\omega\to+\infty$ requires $\tau>-\beta_f/2$ and at $\omega\to-\infty$ requires $\tau<\beta_f/2$, so $G^{\rm sc}_0(it)$ is analytic
in the strip $|{\rm Im}\,t|<\beta_f/2$. 

From (\ref{eq:betaf}) we see that, for integer $\Delta$, 
$G^{\rm sc}_0(\tau)$ is periodic in $\tau$ with period $\beta_f$,
and for $\Delta $ half-integer it is antiperiodic. Taking the operators at two real times
$G^{\rm sc}_0(it_2-it_1) = \Tr\left(V(it_2)V(it_1)\right)$,
we see that on the fake thermal circle the two operators
are shifted by $\beta_f/2$, i.e.\ they are at antipodal locations.
This is why the correlator is nonsingular when the two operator
insertions are at coincident real times.

To summarize, in the semiclassical regime the DSSYK $\beta = 0$
correlator $\Tr\left(V(\tau_2 )V(\tau_1 )\right)$
behaves like a fake temperature thermal correlator
\begin{align}
\Tr\left({\rm e}^{-\beta_fH}V(\tau_2 + \beta_f/2 )V(\tau_1 )\right)
=
\Tr\left({\rm e}^{-\beta_fH/2}V(\tau_2){\rm e}^{-\beta_fH/2}V(\tau_1 )\right)
\end{align}
In the DSSYK/de Sitter proposals of
~\cite{susskind2021entanglementchaossitterholography,
susskind2022scramblingdoublescaledsyksitter,
lin2022infinitetemperatureshot,
rahman2024dsjtgravitydoublescaled,
susskind2022sitterspacedoublescaledsyk,
narovlansky2023doublescaledsyksitterholography}
it is this fake
temperature that plays the role of the de Sitter temperature.

The fake temperature analysis above was, strictly speaking,
only valid in the $\lambda\to 0$ semiclassical limit. Indeed for finite
$\lambda$ the DSSYK $\beta = 0$
correlator $\Tr\left(V(\tau_2 )V(\tau_1 )\right)$
can be evaluated exactly~\cite{Berkooz:2018jqr}, and has
no Euclidean time periodicity in $\tau_2-\tau_1$; the quantum corrections also smooth out the singularity of (\ref{eq:betaf}) at $\tau_2-\tau_1 = \pm\beta_f/2$, the coincident point on the fake circle, since the exact $\beta=0$ correlator $\sum_{ij}|V_{ij}|^2 e^{\tau(E_i-E_j)}/D$ is an entire function of $\tau$ for a bounded spectrum.
So it is worth investigating to what extent the fake thermal circle
picture survives at all for small but finite $\lambda$.

We notice first from (\ref{eq:matchJ}) that,
beyond leading order in finite $\lambda$,
the value of $\mathcal J$ depends on $\Delta$ and thus
the operator size/dimension.
With this caveat, we can
write an expression for the ``one-loop" finite $\lambda$
corrections to the fake-thermal two-point function:
\begin{equation}
G_0(it)=\cosh(\mathcal Jt)^{-2\Delta}\Big[1+\lambda\,g(\mathcal Jt)+O(\lambda^2)\Big],\qquad \mathcal J^2=\frac{1-q^\Delta}{\Delta},
\end{equation}
where $g(\mathcal Jt)$ is the one-loop correction.
Taking now $\Delta = 1$, and
using the exact DSSYK correlator, we can produce an approximate
closed form expression, valid for $\mathcal Jt \gtrsim 1.5$ in the
limit $\lambda\to0$:
\begin{equation}\label{eq:loopg}
g(\mathcal Jt) \simeq -\frac{3}{4}({\mathcal Jt} - 1)
\end{equation}
This shows that the leading order effect at these times is, up to
an overall constant, to
renormalize the coupling that sets the periodicity, rather than
to spoil the periodicity. Denoting this renormalized coupling by $\Jf$,
and combining (\ref{eq:matchJ}) and (\ref{eq:loopg}), we get
\begin{align}\label{eq:Jexpand}
\Jf &\simeq \mathcal J\Big(1+\frac{3}{8}\lambda\Big) = \sqrt\lambda\Big(1-\frac{1}{4}\lambda\Big)\Big(1+\frac{3}{8}\lambda\Big)  = \sqrt{\lambda}\Big(1 + \frac{\lambda}{8}\Big)\\
\cosh(\mathcal J t)^{-2}\big[1+\lambda g(\mathcal J t)\big] &\simeq
\Big(1+\frac{3\lambda}{4}\Big)\cosh\!\big(\Jf\, t\big)^{-2}
\nonumber
\end{align}
and correspondingly $\betaf = \pi/\Jf$.

We conclude that the 
fake-thermal analysis of the DSSYK two-point function 
is consistent for finite but
small $\lambda$ provided we restrict to sufficiently small
times $t$ such that
\begin{equation}\label{eq:gcheck}
\lambda\vert g({\mathcal Jt}) \vert \simeq
\frac{3}{4}\lambda(\sqrt{\lambda} t - 1) \ll 1
\end{equation}
In the results below we satisfy
this constraint out to twice the scrambling time: 
at $t=2t_*$ the left side of (\ref{eq:gcheck}) is $0.15$,
$0.064$ for $q=0.9$, $0.975$ respectively.


\section{How to fold in DSSYK}\label{sec:fold}

Consider the following $\beta = 0$ Euclidean OTOC
in the DSSYK model:
\begin{align}\label{eq:stdO}
\Tr \left( 
V(\tau_4)W(\tau_3)V(\tau_2)W(\tau_1)\,\right)
\end{align}
where for simplicity we will assume $\Delta_V$ = $\Delta_W$ = 1.

We can rewrite this OTOC in a form natural for the chord picture where the
OTOC is computed from the action of the transfer matrix moving around
the fake thermal circle:
\begin{align}\label{eq:gapdef}
\Tr \left(\e^{-g_aH} 
V(0)\e^{-g_bH} W(0) \e^{-g_cH} V(0) \e^{-g_dH} W(0)\,\right)
\end{align}
where the ``gaps" are given by:
\begin{align}\label{eq:gapdefb}
(g_a,\, g_b,\, g_c,\, g_d)
= (\,\tau_{14} +\delta_a, \tau_{43} + \delta_b,
\tau_{32} + \delta_c, \tau_{21} + \delta_d \,)
\end{align}
where $\tau_{14} = \tau_1 - \tau_4$ etc.\ and the $\delta_i$
are additional
contributions to the gaps in analogy to our fake thermal analysis
of the two-point function.
The disconnected part of this 4-point function is the product
of $\Tr(V(\tau_4)V(\tau_2))$ and $\Tr(W(\tau_3)W(\tau_1))$.
Consistency with our fake thermal analysis of the $VV$ two-point
function requires the two $V$'s to be antipodal on the fake circle,
$\delta_b + \delta_c = \beta_f/2$ (equivalently $\delta_d+\delta_a=\betaf/2$),
and likewise for the $WW$ two-point function,
$\delta_c + \delta_d = \beta_f/2$ (equivalently $\delta_a+\delta_b=\betaf/2$).
These conditions fix $\delta_a=\delta_c$, $\delta_b=\delta_d$ and
$\delta_a+\delta_b=\betaf/2$, leaving a one-parameter family.
The symmetric choice
$\delta_a$$=$$\delta_b$$=$$\delta_c$$=$$\delta_d$$=$$\beta_f/4$
is singled out by the symmetry between $V$ and $W$.

Let's apply this now to the following $\beta = 0$
standard connected DSSYK OTOC:
\begin{align}\label{eq:fstd}
F_{\rm std}(t) = \Tr \left( 
V(it)W(0)V(it)W(0)\,\right)
\end{align}
With our gap mapping to the fake thermal circle this becomes:
\begin{align}\label{eq:Fstd}
F_{\rm std}(t) &= \Tr \left( {\rm e}^{-\betaf H/4}\,
V(it){\rm e}^{-\betaf H/4}\,W(0){\rm e}^{-\betaf H/4}\,V(it)
{\rm e}^{-\betaf H/4}\,W(0)\,\right) \nonumber\\
&= \Tr \left( {\rm e}^{-\betaf H}\,
V(\frac{3\betaf}{4}+it)W(\frac{\betaf}{2})V(\frac{\betaf}{4}+it)W(0)\,\right)
\end{align}
where in the second expression we see that $F_{\rm std}(t)$
looks like a standard thermal OTOC at inverse temperature
$\beta_f$ after applying the regulator 
prescription of Maldacena, Stanford, and Shenker (MSS)~\cite{maldacena2016bound}.


\begin{figure}[t]
\centering
\begin{tikzpicture}[>=Stealth,
  dot/.style={circle,fill,inner sep=1.7pt},
  midarrow/.style={postaction={decorate,decoration={markings,mark=at position 0.5 with {\arrow{>}}}}},
  fwd/.style={very thick,blue!65!black,midarrow},
  bwd/.style={very thick,red!85!black,dashed,midarrow},
  lab/.style={font=\small}]
\begin{scope}[xshift=-3.9cm]
  \draw[gray!35] (0,0) circle (1.6);
  \draw[bwd] (90:1.6) -- (90:2.0) arc (90:180:2.0) -- (180:1.6);
  \draw[fwd] (180:1.6) arc (180:-90:1.6);
  \draw[bwd] (-90:1.6) -- (-90:2.0) arc (-90:0:2.0) -- (0:1.6);
  \draw[fwd] (0:1.6) -- (0:1.3) arc (0:-270:1.3) -- (90:1.6);
  \node[dot,blue!65!black] at (90:1.6) {};   \node[lab,above=3pt,blue!65!black] at (90:2.0) {$W(\tau_1)$};
  \node[dot,red!85!black]  at (180:1.6) {};  \node[lab,left=3pt,red!85!black]   at (180:2.0) {$V(\tau_2)$};
  \node[dot,blue!65!black] at (-90:1.6) {};  \node[lab,below left=1pt,blue!65!black] at (-100:1.9) {$W(\tau_3)$};
  \node[dot,red!85!black]  at (0:1.6)  {};   \node[lab,right=3pt,red!85!black]  at (0:2.0)  {$V(\tau_4)$};
  \node[lab,red!85!black]  at (135:2.75) {$-\tfrac{\beta_f}{4}-it$};
  \node[lab,red!85!black]  at (-45:2.75) {$-\tfrac{\beta_f}{4}-it$};
  \node[lab,blue!65!black] at (45:1.0)   {$\tfrac{3\beta_f}{4}+it$};
  \node[lab,blue!65!black] at (-135:0.65) {$\tfrac{3\beta_f}{4}+it$};
\end{scope}
\begin{scope}[xshift=3.9cm]
  \draw[gray!35] (0,0) circle (1.6);
  \draw[fwd] (90:1.6) arc (90:-180:1.6);
  \draw[bwd] (180:1.6) -- (180:2.0) arc (180:270:2.0) -- (270:1.6);
  \draw[fwd] (-90:1.6) -- (-90:1.3) arc (-90:-360:1.3) -- (0:1.6);
  \draw[bwd] (0:1.6) -- (0:2.0) arc (0:90:2.0) -- (90:1.6);
  \node[dot,blue!65!black] at (90:1.6) {};   \node[lab,above=3pt,blue!65!black] at (90:2.0) {$W(\tau_1)$};
  \node[dot,red!85!black]  at (180:1.6) {};  \node[lab,left=3pt,red!85!black]   at (180:2.0) {$V(\tau_2)$};
  \node[dot,blue!65!black] at (-90:1.6) {};  \node[lab,below right=1pt,blue!65!black] at (-80:1.9) {$W(\tau_3)$};
  \node[dot,red!85!black]  at (0:1.6)  {};   \node[lab,right=3pt,red!85!black]  at (0:2.0)  {$V(\tau_4)$};
  \node[lab,red!85!black]  at (225:2.75) {$-\tfrac{\beta_f}{4}+it$};
  \node[lab,red!85!black]  at (45:2.75)  {$-\tfrac{\beta_f}{4}+it$};
  \node[lab,blue!65!black] at (-55:0.90)  {$\tfrac{3\beta_f}{4}-it$};
  \node[lab,blue!65!black] at (135:0.55) {$\tfrac{3\beta_f}{4}-it$};
\end{scope}
\end{tikzpicture}
\caption{
Illustration of the  $n$$=$$0$ (left) and $n$$=$$1$ (right) folds of the fake-MSS configuration
on the fake thermal circle. Solid arcs are forward Euclidean gaps, dashed arcs are backward ones; Euclidean time increases clockwise and the real-time displacements of the $V$'s are suppressed. 
The $n$$=$$0$ folding prescription
moves each $V$ backwards by $\beta_f/2$ along the circle (and by $-2it$ in real time),
to fake positions $-\betaf/4$ and $\betaf/4$, while the operator ordering in the
trace is unchanged. The Euclidean contour is therefore folded:
starting from $W(\tau_1)$ it runs backwards by $\betaf/4$ (dashed) to $V(\tau_2)$,
forwards by $3\betaf/4$ to $W(\tau_3)$, backwards by $\betaf/4$ to $V(\tau_4)$, and forwards by
$3\betaf/4$ back to $W(\tau_1)$.
After the common $\betaf/4$ offsets are removed (eq.~(\ref{eq:Ffold0})), the $n$$=$$1$ gaps are the $n$$=$$0$ gaps with every sign reversed and $t\to-t$. Reversing each gap is a symmetry of the $\beta=0$ chord expansion (non-vanishing chord diagrams contain an even number of $H$'s), so that
$F_{\rm fold}^{(1)}(t)$ = $F_{\rm fold}^{(0)}(-t)$ = $(F_{\rm fold}^{(0)}(t))^*$ for any $\Delta_V,\Delta_W$;
the choice of $n$ determines the side from which the continued eikonal amplitude approaches its branch cut.}
\label{fig:circle}
\end{figure}
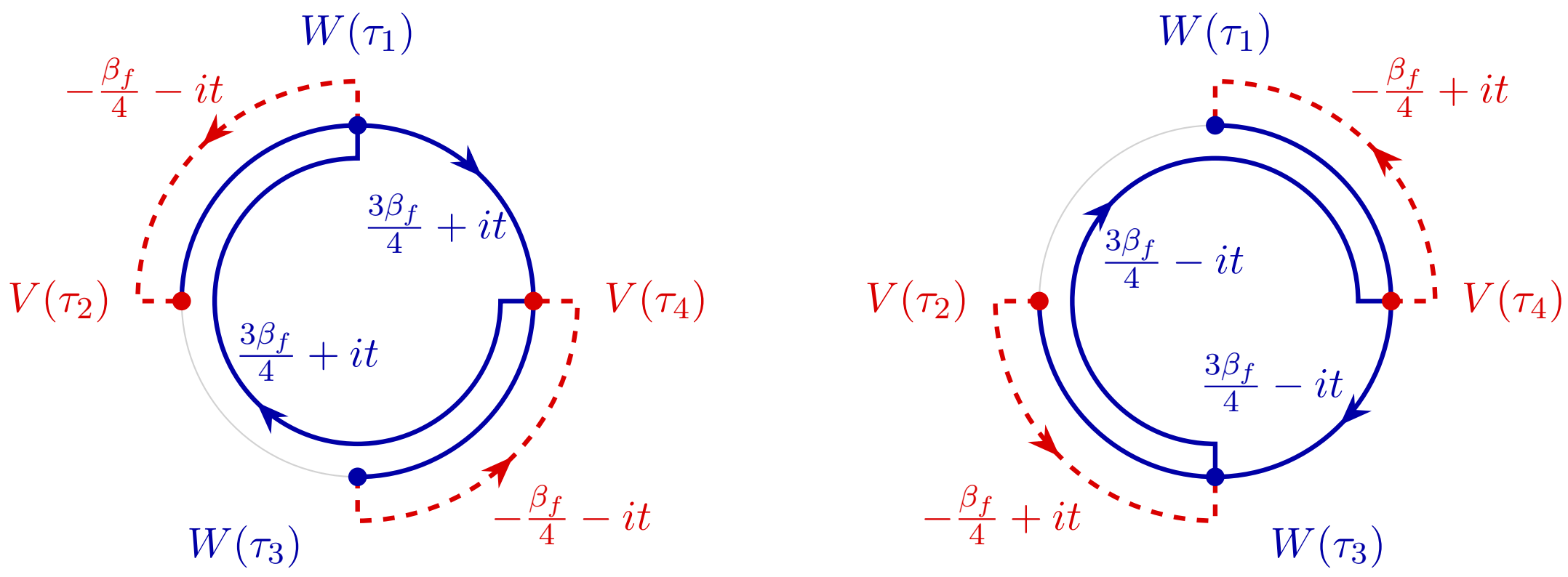


Now we apply the folding prescription
of~\cite{harlow2026antiscramblingeuclideanfoldsobserver},
the $n$$=$$0$ version of (\ref{eq:HZ}), 
using the fake temperature instead of the
thermal temperature.
We call the result the folded OTOC
$F_{\rm fold}(t)$. The HZ prescription $\tau_1\to -\tau_1$,
$\tau_2\to - \tau_2$, $\tau_3\to \betaf - \tau_3$,
and $\tau_4\to \betaf - \tau_4$, 
applied to the
second expression in (\ref{eq:Fstd}), gives:
\begin{align}\label{eq:Ffoldfake}
F_{\rm fold}(t) = \Tr \left( {\rm e}^{-\betaf H}\,
V(\frac{\betaf}{4}-it)W(\frac{\betaf}{2})V(-\frac{\betaf}{4}-it)W(0)\,\right)
\end{align}
The four Euclidean gaps are now $(3\betaf/4+it,\,-\betaf/4-it,\,3\betaf/4+it,\,-\betaf/4-it)$: two of them are negative, which is the fold.
We see that, in the fake thermal circle picture, the folded OTOC is obtained from the standard OTOC
by letting $t\to -t$ and moving each $V$ backwards by half the fake
circle, so that the operator string is no longer ordered in fake
Euclidean time; this is illustrated
in the left plot of Figure~\ref{fig:circle}.
The equivalent folding with the $n$$=$$1$ prescription is
shown in the right plot of Figure~\ref{fig:circle}.
Since $F_{\rm fold}^{(1)}(t)$ = $(F_{\rm fold}^{(0)}(t))^*$,
we will confine our analysis to the $n$$=$$0$ case.

We can rewrite $F_{\rm fold}(t)$ as:
\begin{align}\label{eq:Ffoldgaps}
F_{\rm fold}(t) = \Tr \left( {\rm e}^{-\betaf H/4}\,
V(\frac{-\betaf}{2}-it)\,
{\rm e}^{-\betaf H/4}\,W(0)\,
{\rm e}^{-\betaf H/4}\,V(\frac{-\betaf}{2}-it)\,
{\rm e}^{-\betaf H/4}\,W(0)\,\right)
\end{align}
To regain a $\beta=0$ DSSYK OTOC we remove
all the $\beta_f/4$ gaps to obtain:
\begin{align}\label{eq:Ffold0}
F_{\rm fold}(t)
= \Tr\left(
V(\frac{-\betaf}{2}-it)W(0)V(\frac{-\betaf}{2}-it)W(0)\,\right)
\end{align}
As noted by \cite{harlow2026antiscramblingeuclideanfoldsobserver},
in the folded OTOC the roles of $V$ and $W$
have switched, since $V$ acts before $W$ rather than the other way around
as was the case for $F_{\rm std}(t)$. In the present case this
switch is equivalent to $t\to -t$. To see this, note that
$f(\tau)\equiv\Tr(V(\tau)WV(\tau)W)$ is even in $\tau$ for any matter
content: expanding $V(\tau)=e^{\tau H}Ve^{-\tau H}$, every term in the
disorder average is a chord diagram in which the $H$'s are paired, so
only even total powers of $\tau$ survive. (Equivalently, conjugation by
$(-1)^{\hat n}$ sends $T\to-T$ while commuting with the matter chord
operations.) Hence $F_{\rm fold}(t)=f(-\tfrac{\betaf}{2}-it)=f(\tfrac{\betaf}{2}+it)$, and since $f$ has real
Taylor coefficients, $f(\bar z)=f(z)^*$, so that
\begin{equation}\label{eq:conj}
F_{\rm fold}(-t) = f\big(\tfrac{\betaf}{2}-it\big) = F_{\rm fold}(t)^* ,
\end{equation}
i.e.\ ${\rm Re}\,F_{\rm fold}$ is even and ${\rm Im}\,F_{\rm fold}$ is
odd in $t$. Below we will always look at $F_{\rm fold}(t)$ for $t>0$.
The $n=1$ fold is $f(\tfrac{\betaf}{2}-it)$: its gaps are those of the
$n=0$ fold with all signs reversed and $t\to-t$. By the same parity it
equals $F_{\rm fold}(-t)=F_{\rm fold}(t)^*$.

Two remarks are in order. First, (\ref{eq:Ffold0}) is an ordinary
$\beta=0$ correlator of the DSSYK model, but one in which the $V$
operators have been evolved backwards in Euclidean time by half the
fake period. This is well-defined because the DSSYK energy spectrum is
bounded, the condition required
in~\cite{harlow2026antiscramblingeuclideanfoldsobserver}.
Second, the fold displaces each $V$ from the $W$'s by $\betaf/2$ in
Euclidean time, the half-width of the strip $|{\rm Im}\,t|<\betaf/2$ in
which the semiclassical two-point function (\ref{eq:betaf}) is
analytic. The folded operator
$\tilde V=e^{-\betaf H/2}Ve^{\betaf H/2}$ is therefore far from bounded
in the semiclassical sense: $\Tr(\tilde V^\dagger\tilde V)=G_0(\betaf)$ is exponentially large in $1/\lambda$.
The OTOC itself is however well-behaved: we will show in 
the next section that its
spectral weight falls off fast enough for the folded integral to
converge with room to spare.  It is still not obvious {\it a priori} that
the semiclassical eikonal formula should match the exact correlator at
the fold. That it does, at the
percent level (Section~\ref{sec:numerics}), shows that we satisfy
the criterion of~\cite{harlow2026antiscramblingeuclideanfoldsobserver}
that the continuation should commute with the
semiclassical limit.
The large norm of $\tilde V$ makes the numerical
evaluation quite delicate: the intermediate states in the chord-space
evaluation of (\ref{eq:Ffold0}) have norms growing like
$e^{{\rm const}/\lambda}$ even though the result is $O(1)$.


\section{Long-time limit of the exact DSSYK OTOC}\label{sec:longtime}

Berkooz--Isachenkov--Narovlansky--Torrents~\cite{Berkooz:2018jqr}
derived an exact expression for the standard OTOC four-point function;
using the notation of 
Mukhametzhanov~\cite{mukhametzhanov2023largepsykchord} 
and introducing four Lorentzian times $u_1$,$u_2$,$u_3$,$u_4$, we 
can write an exact $\beta = 0$ expression for (\ref{eq:stdO}):
\begin{align}\label{eq:exactOTOC}
\Tr \left( 
V(iu_1)W(iu_2)V(iu_3)W(iu_4)\,\right) &= \\
&\!\!\!\!\!\!\!\!\!\!\!\!\!\!\!\!\!
\!\!\!\!\!\!\!\!\!\!\!\!\!\!\!\!\!\!\!\!\!
\int_0^{\pi/\lambda}\,\prod_{i=1}^4 \left( ds_i\,
\rho_q(s_i) \right)
{\rm e}^{-(g_a E_1 + g_b E_2 + g_c E_3 + g_d E_4)}  \left( \Gamma_{12} \Gamma_{23}
\Gamma_{34} \Gamma_{41} \right)^{1/2}\,
\begin{Bmatrix}
  \Delta & s_1 & s_2 \\
  \Delta & s_3 & s_4
\end{Bmatrix} \nonumber
\end{align}
where we have used the gap notation of (\ref{eq:gapdef}),(\ref{eq:gapdefb})
with
$(g_a,\, g_b,\, g_c,\, g_d)
= (iu_{41},iu_{12},iu_{23},iu_{34})$, $u_{ij} = u_i-u_j$.
We also have:
\begin{align}
\rho_q(s) &= \frac{1}{2\pi \Gamma_q(2is) \Gamma_q(-2is)}
\\
\Gamma_{ij} &= \left( \frac{\Gamma_q(\Delta + i(s_i+s_j))
\Gamma_q(\Delta + i(s_i-s_j))\Gamma_q(\Delta - i(s_i-s_j))
\Gamma_q(\Delta - i(s_i+s_j))}{\Gamma_q(2\Delta)} \right) \nonumber
\end{align}
where $\rho_q(s)$ is the density of states, the $\Gamma_q$ are q-Gamma functions, and the curly bracket
object in (\ref{eq:exactOTOC}) is a q-6j symbol.
The $s_i$ parameters come from using the DSSYK spectral 
formula (\ref{eq:energy})
in our units with $J=1$.

To obtain $F_{\rm std}(t)$ we should use the gaps
$(-it, it, -it, it)$, corresponding to $u_1 = u_3 = t$,
$u_2 = u_4 = 0$. The folded OTOC (\ref{eq:Ffold0}) corresponds to the gaps
$(\beta_f/2 +it, -\beta_f/2 - it, \beta_f/2 +it,-\beta_f/2 - it)$,
i.e.\ $u_1 = u_3 = -t + i\beta_f/2$, $u_2 = u_4 = 0$.
Thus $F_{\rm fold}(t)$ is computed from the same integrand
as $F_{\rm std}(-t)=F_{\rm std}(t)$, but with an additional factor
e$^{-(\beta_f/2)(E_1-E_2+E_3-E_4)}$.

Following the analysis of Mukhametzhanov~\cite{mukhametzhanov2023largepsykchord},
we can simplify the pieces of the integrand in
the spectral representation (\ref{eq:exactOTOC})
by taking the limit $\lambda \to 0$ with $\lambda s$
and $s_i - s_j$ fixed (his eq.~(5.6)), and then the ``long time'' limit
$\lambda\to0$ with $z$ fixed (his eq.~(5.23)), i.e.\
$\lambda_L t - \log(1/\lambda)$ fixed, with $z$ defined below.

The first limit defines the same regime where we
saw that the partition function is dominated by
the saddle (\ref{eq:saddle}).

The 4-dimensional integral for the OTOC in 
(\ref{eq:exactOTOC}) can be rewritten as an integral
over the average energy $s = (s_1+s_2+s_3+s_4)/4$ and integrals over
energy differences. From
Mukhametzhanov's eq.~(5.18) we see that in the unfolded case
the integral over $s$ is dominated by the same saddle
as the partition function.

Now consider a saddle analysis in the $\beta $$=$$0$ folded case.
For simplicity of the discussion let's enforce $s_1$$=$$s_3$,
$s_2$$=$$s_4$ and look at the behavior of the
saddle in the $s_1$-$s_2$ plane. Denoting the saddle exponent
as $\mathcal S(\theta_1,\theta_2)/\lambda$, where $s_1=\theta_1/\lambda$,
$s_2=\theta_2/\lambda$, we see from
Mukhametzhanov's eqs.~(5.7), (5.11), and (5.17) that the total entropic
contribution to $\mathcal S(\theta_1,\theta_2)$ is
$\tfrac12\big(\mathcal S(\theta_1) + \mathcal S(\theta_2)\big)$, with $\mathcal S(\theta )$ as defined in
(\ref{eq:dstates}).

To understand the contribution from the
alternating folds
first note that since the four folds
add to zero the symmetric saddle configuration $s_1$$=$$s_2$ is not affected by the folds. Looking then at asymmetric configurations
with $s_1$$\neq$$s_2$, we see that the fold
factor e$^{-(\beta_f/2)(E_1-E_2+E_3-E_4)}$
pushes in the direction of $E_1$$<$$E_2$, contributing
$(\beta_f/2)(2\sqrt\lambda)({\rm cos}\,\theta_1-{\rm cos}\,\theta_2
+{\rm cos}\,\theta_3-{\rm cos}\,\theta_4)
= 2\pi({\rm cos}\,\theta_1-{\rm cos}\,\theta_2)$
to $\mathcal S(\theta_1,\theta_2)$,
to leading order in small $\lambda$.

However asymmetric configurations in energy require that the matter
operator insertions are transferring energy from one segment to the
other. Denoting the energy transfer by $\omega$, we already
saw in (\ref{eq:tailsb}) that
the spectral weight in $\omega$ decays like $\e^{-\beta_f|\omega|/2}$.
Using $\omega = E(\theta_1)-E(\theta_2)
\simeq (2/\sqrt\lambda)(\theta_1-\theta_2)$ at the energy midpoint,
this becomes $\e^{-\pi|\theta_1-\theta_2|/\lambda}$ to leading order in small $\lambda$.
Matching to the gamma function asymptotics of
Mukhametzhanov's eqs.~(5.7) and (5.17), we find that the total contribution
of this type to $\mathcal S(\theta_1,\theta_2)$ is
$-3\pi|\theta_1-\theta_2|$: the four vertex factors give
$-2\pi|\theta_1-\theta_2|$, and the factor $\Gamma(i(\nu_2+\nu_4))$ in
the asymptotic form (5.7) of the 6j-symbol gives a further
$-\pi|\theta_1-\theta_2|$ on the slice $s_1$$=$$s_3$, $s_2$$=$$s_4$.
These asymptotic forms are derived at fixed $s_i-s_j$, so they hold for
$\lambda\ll|\theta_1-\theta_2|\ll1$.

All together, the folded OTOC saddle behavior 
in the $s_1$-$s_2$ plane is determined
by maximizing
\begin{align}\label{eq:2Dsaddle}
\mathcal S(\theta_1,\theta_2) = (\pi\theta_1 - \theta_1^2)
+(\pi\theta_2 - \theta_2^2)
+2\pi({\rm cos}\,\theta_1 - {\rm cos}\,\theta_2 )
-3\pi|\theta_1-\theta_2|
\end{align}
The maximum is at $\theta_1 = \theta_2 = \pi/2$, the
$\beta = 0$ saddle point of
(\ref{eq:saddle}). It is not a stationary point of the smooth part of
(\ref{eq:2Dsaddle}), whose gradient there is $(-2\pi,+2\pi)$, but a cusp.
Moving off the diagonal by $d=\theta_2-\theta_1>0$, the fold gain
is $2\pi d$ while the spectral weight cost is $3\pi d$, so
the folded integral converges with a margin. Note that the fold depth could be increased
by a factor of $3/2$ before the cusp is lost; we return to this
observation in Section~\ref{sec:fail}.

In Mukhametzhanov's long-time limit,
eq.~(\ref{eq:exactOTOC}), normalized by the product of
two-point functions, reduces to (eq.~(5.20) of~\cite{mukhametzhanov2023largepsykchord})
(see also~\cite{Maldacena:2016upp}):
\begin{align}\label{eq:Mz}
z^{-2\Delta}\,U(2\Delta,1;1/z)\;;\qquad
z = \frac{\lambda^{3/2}}{2\lambda_L}\,\,
\frac{\cosh\left(\frac{\lambda_L}{2}(u_1-u_2+u_3-u_4)\right)}
{\cosh\left(\frac{\lambda_L}{2}(u_1-u_3)\right)
\cosh\left(\frac{\lambda_L}{2}(u_2-u_4)\right)}
\end{align}
where $U(2\Delta,1;1/z)$ is the confluent hypergeometric function,
and $\lambda_L$ is the Lyapunov exponent; $\lambda_L = 2\sqrt{\lambda}$
at leading order for small $\lambda$. In the scaling limit only the
growing exponential in the numerator survives, and Mukhametzhanov writes
$e^{\frac{\lambda_L}{2}(u_2+u_4-u_1-u_3)}$ in place of $2\cosh(\dots)$
(in his time-ordering conventions); we have symmetrized it to the
$\cosh$, which is equivalent in the scaling limit and makes (\ref{eq:Mz})
even in $t$, as the exact $\beta=0$ correlator is.

For $\Delta=1$, the integral representation $U(a,b;x)=\frac{1}{\Gamma(a)}\int_0^\infty dt\,e^{-xt}t^{a-1}(1+t)^{b-a-1}$ gives
\begin{equation}
U(2,1;\tfrac1z)=\int_0^\infty dt\,e^{-t/z}\frac{t}{(1+t)^2}
\;\xrightarrow{\;t=zp\;}\;
z^{-2}U(2,1;\tfrac1z)=\int_0^\infty dp\,\frac{p\,e^{-p}}{(1+zp)^2},
\end{equation}
and with $\frac{1}{(1+zp)^2}=\int_0^\infty dq\,q\,e^{-q}e^{-zpq}$,
\begin{equation}
F_{\rm std}(t) \to \int_0^\infty\!\!dp\,dq\;pq\,e^{-p-q-zpq}=\Phi(z),\qquad\text{i.e.\ } z\equiv a 
\end{equation}
This is the eikonal function of (\ref{eq:Phi}),
with 
\begin{equation}\label{eq:zcosh}
a(t) = z = \frac{\lambda^{3/2}}{2\lambda_L}\,{\rm cosh}(\lambda_L t)
\to \frac{\lambda}{4}\,{\rm cosh}(2\sqrt{\lambda}t)
\end{equation}


\begin{figure}[htb]
\begin{center}
\includegraphics[scale=.99]{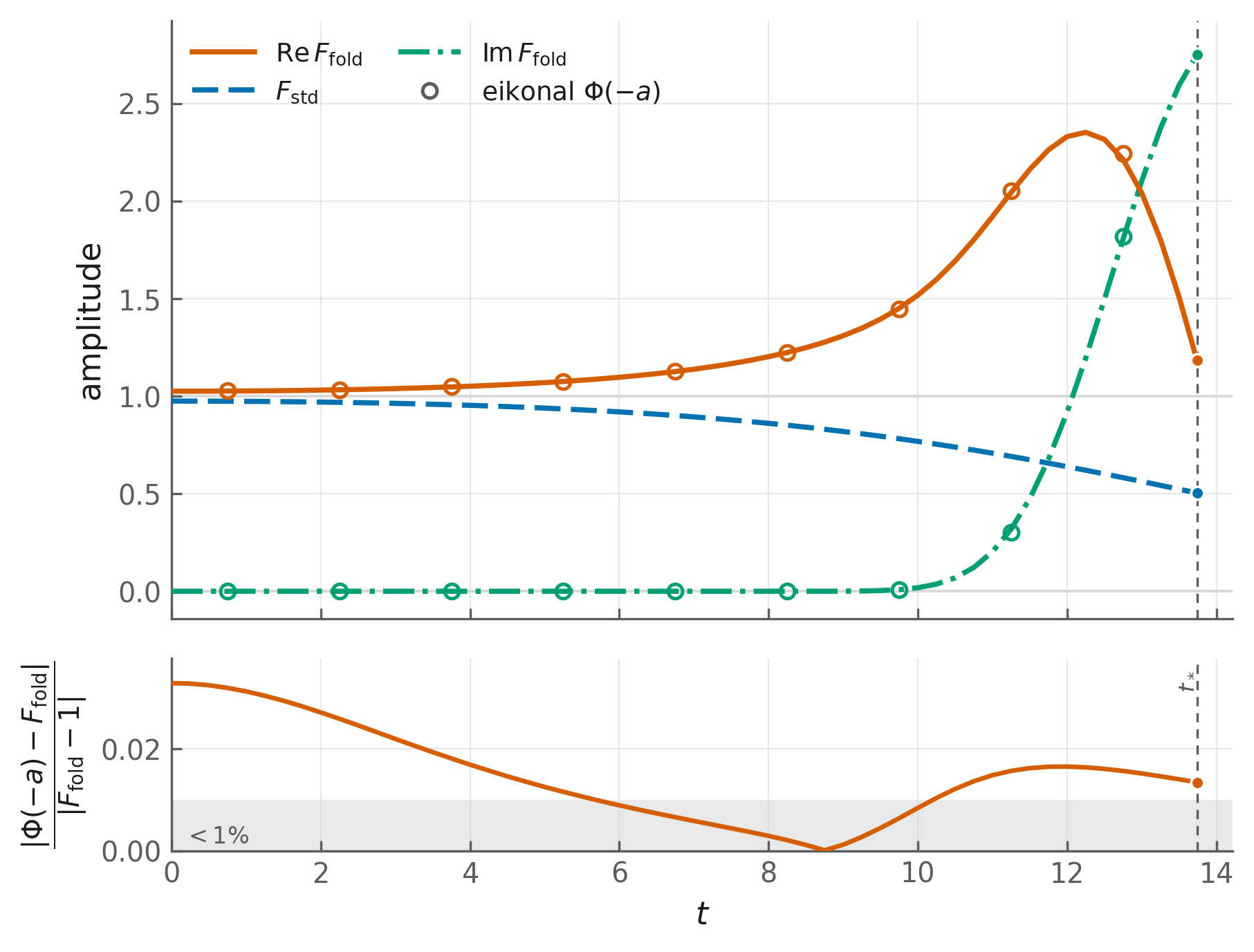}
\caption{Plots of the standard OTOC $F_{\rm std}(t)$ and the real and imaginary
parts of the folded OTOC $F_{\rm fold}(t)$ in the exact DSSYK model,
for $q = 0.975$, implying $\lambda = 0.0253$, $\beta_f \simeq \pi/\sqrt\lambda = 19.7$,
and scrambling time $t_* = 13.74$. The lower figure shows the 
relative deviation of the complex amplitudes,
$\delta = |\Phi(-a(t))-F_{\rm fold}(t)|/|F_{\rm fold}(t)-1|$ (the $\delta$ column of Table~\ref{tab:q975}),
where $a(t)$ was computed by matching $\Phi(a(t))$ to
$F_{\rm std}(t)$.
}
\label{fig:OTOC975}
\end{center}
\end{figure}


\subsection{The fold as a rotation of \texorpdfstring{$z$}{z}}\label{sec:rotation}
Since $z$ depends on the times only through
$\cosh\left(\frac{\lambda_L}{2}(u_1-u_2+u_3-u_4)\right)$
(the two $\cosh$'s in the denominator of (\ref{eq:Mz}) are unaffected,
because $u_1-u_3=u_2-u_4=0$),
the folding process, $u_1 = u_3 = -t \to -t + i\betaf/2$, takes
\begin{equation}\label{eq:zrot}
z\to z\,\frac{\cosh(\lambda_Lt_\mathrm{eff}+i\lambda_L\betaf/2)}{\cosh \lambda_Lt_\mathrm{eff}}\;,
\qquad t_\mathrm{eff} = \frac{1}{2}(u_1-u_2+u_3-u_4)\,.
\end{equation}
To leading order in small $\lambda$, $\lambda_L\beta_f/2=\pi$, so
$z_{\rm fold}=-z$. The side from which the negative axis is approached
follows from performing the Euclidean shift continuously: writing
$u_1=u_3=-t+i\varphi/2$ with $\varphi$ increasing from $0$ to $\betaf$,
(\ref{eq:Mz}) gives
$z(\varphi)\propto\cosh(-\lambda_Lt+i\lambda_L\varphi/2)
=\cosh(\lambda_Lt)\cos(\lambda_L\varphi/2)-i\sinh(\lambda_Lt)\sin(\lambda_L\varphi/2)$,
whose imaginary part is negative for $t>0$ and $0<\lambda_L\varphi/2<\pi$.
The contour thus approaches the negative axis through the lower half of the $z$-plane:
\begin{equation}
z_{\rm fold}=-z-i0\qquad\Longrightarrow\qquad
F_{\rm fold}(t)\;\to\;\Phi(-a-i0)\qquad (t>0),
\end{equation}
while for $t<0$ the approach is from above, consistent with (\ref{eq:conj}).
With $w=1/a$ and $\Phi(a)=w^2\big[(1+w)e^wE_1(w)-1\big]$, the fold sends
\begin{equation}
1/a\to  \frac{1}{-a-i0} = -w+i0
\end{equation}
and on the
principal branch $E_1(-w+i0)=-\mathrm{Ei}(w)-i\pi$, hence
\begin{align}\label{eq:foldPhi}
\Phi(-a-i0)&=w^2\Big[(1-w)\,e^{-w}\big(-\mathrm{Ei}(w)-i\pi\big)-1\Big]\simeq1+4a+18a^2+\dots ,\nonumber\\
\mathrm{Im}\,\Phi(-a-i0)&=\pi w^2(w-1)e^{-w}
\end{align}
using $\mathrm{Ei}(w)\simeq\frac{e^w}{w}\big(1+\frac1w+\frac{2!}{w^2}+\frac{3!}{w^3}+\dots\big)$.


\subsection{The thermal fold at high temperatures}\label{sec:fail}
We have examined the fold in the thermal state of DSSYK at high temperatures: $\beta$ nonzero but  $\ll O(1/\lambda^{3/2})$, which
is where the Schwarzian regime takes over.
At finite $\beta$ the Lyapunov exponent is $\lambda_L(\beta)=2\pi v/\beta$,
with $v$ the solution of (\ref{eq:saddle})~\cite{Maldacena:2016hyu}, and the rotation of $z$ by $\pi$ in Section~\ref{sec:rotation} requires a
Euclidean shift of $\pi/\lambda_L=\beta/(2v)$; this is the fold depth
that reaches the $a\to-a$ sheet. Equivalently, at finite $\beta$ the
fake inverse temperature is $2\pi/\lambda_L=\beta/v$, and the
prescription is again to fold by half the fake period.
For $\beta \to 0$ we see from the saddle point equation
(\ref{eq:saddle}) that $v\to \beta/\beta_f$ to leading order
in small $\lambda$, so the finite $\beta$
prescription reduces to folding by $\beta_f/2$ in the
$\beta \to 0$ limit.

Explicitly, starting from the MSS-symmetric thermal configuration and
shifting each $V$ by a fold depth $s_d$, the finite-$\beta$ folded OTOC is
\begin{align}\label{eq:Ffoldbeta}
F_{\rm fold}(t) &= \frac{1}{\mathcal N_\beta}\,\Tr \left(\e^{-g_aH}\,V\,\e^{-g_bH}\,W\,\e^{-g_cH}\,V\,\e^{-g_dH}\,W\,\right) \\
(g_a,\, g_b,\, g_c,\, g_d) &= \big(\tfrac{\beta}{4}+s_d+it,\ \tfrac{\beta}{4}-s_d-it,\ \tfrac{\beta}{4}+s_d+it,\ \tfrac{\beta}{4}-s_d-it\big) \nonumber
\end{align}
which reduces to (\ref{eq:Ffold0}) for $\beta\to0$ with $s_d=\betaf/2$.
Here $\mathcal N_\beta=Z(\beta)\,G_V(\beta/2)\,G_W(\beta/2)$ normalizes by
the partition function and by the two thermal two-point functions, which
do not depend on $s_d$ or $t$ because the two $V$'s, and the two $W$'s,
remain separated by $\beta/2$.

For finite $\beta$, we can repeat the saddle point analysis
of the folded OTOC in the $s_1$-$s_2$ plane.
Instead of (\ref{eq:2Dsaddle}) we get the following
expression:
\begin{align}\label{eq:2Dsaddleb}
\mathcal S(\theta_1,\theta_2) =& (\pi\theta_1 - \theta_1^2)
+(\pi\theta_2 - \theta_2^2)\\
&+\sqrt\lambda (\beta+4\frac{\beta}{2v})\,{\rm cos}\,\theta_1
+\sqrt\lambda(\beta-4\frac{\beta}{2v})\,{\rm cos}\,\theta_2
-3\pi|\theta_1-\theta_2| \nonumber
\end{align}
where the Boltzmann terms couple only to the symmetric
combination, while the fold couples only to the
antisymmetric combination.

For small enough $\beta$, there is a 2-dimensional symmetric saddle
$\theta_1$$=$$\theta_2 = (\pi/2)(1-v)$, with $v$ the solution of
(\ref{eq:saddle}), corresponding to a saddle point energy
below the midpoint. With the fold depth $\beta/(2v)$, the slope of
(\ref{eq:2Dsaddleb}) off the diagonal at this saddle is
$2\pi-3\pi=-\pi$ for every $\beta$, by (\ref{eq:saddle}), exactly as at
$\beta=0$: the symmetric saddle remains a cusp maximum and never becomes
locally unstable. Any failure has to come from elsewhere in the
$\theta_1$-$\theta_2$ plane. Close to the midpoint
we can write $\theta_2 = \pi/2 + \epsilon$, $\epsilon$ small,
and observe that, for $\theta_2>\theta_1$:
\begin{align}
\frac{\partial \mathcal S(\theta_1,\theta_2)}{\partial \theta_2}
\simeq -2\epsilon +4\sqrt\lambda\,b -3\pi
\end{align}
where we have defined $b = \beta/(2v) - \beta/4$ as
the effective backwards fold (more generally $b=s_d-\beta/4$ for a fold
depth $s_d$). For $b > 3\pi/(4\sqrt\lambda )$,
corresponding to backward segments 1.5 times larger than
the $\beta = 0$ fold $\beta_f/2$, (\ref{eq:2Dsaddleb}) therefore has a
second, real stationary point in $\theta_2$ near the midpoint, away from
the symmetric saddle; for smaller $b$ the corresponding stationary
points form a complex conjugate pair. Since this uses the form
$-3\pi|\theta_1-\theta_2|$ at $|\theta_1-\theta_2|=O(1)$, outside the
regime in which it was derived, we take it only as an estimate of the
threshold.
We will confirm this threshold with our exact numerical
results in the next section, and show that it
corresponds to a pair of complex conjugate
saddles emerging as dominant over the original
symmetric saddle.

\subsection{The scrambling time}
To define the scrambling time $t_*$, we use the
criterion of~\cite{harlow2026antiscramblingeuclideanfoldsobserver} 
that it is the time when the leading connected
part of the OTOC equals the disconnected part. Using the
expansion of $\Phi(-a(t))$ in (\ref{eq:foldPhi}) this implies
$a(t_*) = 1/4$, and thus
\begin{equation}\label{eq:tscram}
t_* \simeq \frac{1}{2\sqrt{\lambda}}\,\mathrm{log}\frac{2}{\lambda}
\end{equation}
To leading order in small $\lambda$ this
has the Harlow-Zhao form $t_* = (\beta_f/2\pi)\log S_{\rm dS}$ (their eq.~(2.8)),
with $S_{\rm dS}$$=$$2/\lambda = N/p^2$ playing the role of the de Sitter entropy, and $\beta_f$ the inverse temperature.
Although $t_*$ is defined from the leading term of the asymptotic
series, at $a=1/4$ that series is useless (its first five terms sum
to $7.0$, versus $\Phi(-1/4-i0) = 1.26 + 2.76\,i$); all comparisons
below use the closed form (\ref{eq:foldPhi}). It is also worth noting
that $|F_{\rm fold}(t)|$ peaks very close (within about a percent) to the
scrambling time as defined by (\ref{eq:tscram}), so this provides
an alternate definition of the 
scrambling time self-contained within our analysis.
With this estimate the reader can check that the numerical examples
in the next section
easily satisfy the consistency condition (\ref{eq:gcheck}) out to 
twice the scrambling time.


\begin{figure}[htb]
\begin{center}
\includegraphics[scale=.73]{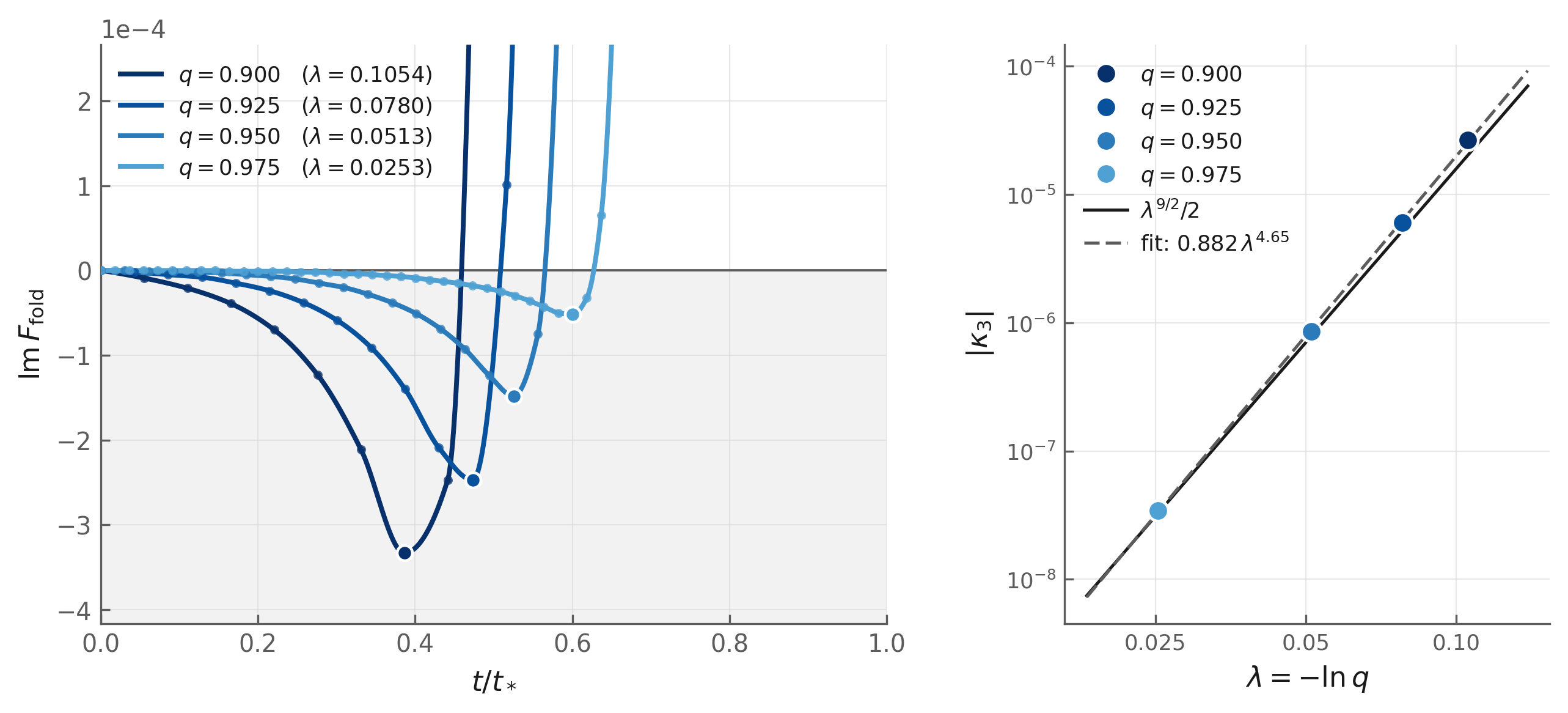}
\caption{Plot of the imaginary
part of the folded OTOC $F_{\rm fold}(t)$ in the exact DSSYK model,
for $q =$ 0.9, 0.925, 0.95, and 0.975, implying 
$\lambda =$ 0.1054, 0.078, 0.0513, and 0.0253. The left plot shows 
the very small negative imaginary part that manifests at early times
from folding using $\beta_f$ computed to NNLO in $\lambda$;
this transitions to a much larger positive imaginary part closer
to the scrambling time. The right plot shows a fit 
using the exact $\beta_f$ to the non-eikonal early time
behavior of Im$\,F_{\rm fold}(t)$. 
The fit is consistent with
Im$\,F_{\rm fold}(t)-{\rm Im}\,\Phi(-a-i0)$ $\simeq -\kappa_3\,t^3$,
with $\kappa_3 = \lambda^{9/2}/2$.
}
\label{fig:OTOC_artifact}
\end{center}
\end{figure}


\section{Numerical analysis of the folded OTOC}\label{sec:numerics}

We compute $F_{\rm std}(t)$ and $F_{\rm fold}(t)$
numerically in the exact DSSYK model.
We evaluate (\ref{eq:fstd}) and (\ref{eq:Ffold0}) directly in the
chord Hilbert space with two matter chords, acting with the transfer
matrix and the matter operators in sequence on the empty state; the
states are labeled by the numbers of $H$-chords in the regions
separated by the open matter chords, $|n_0,n_1,n_2\rangle$, with
$n_0+n_1+n_2\le n_{\rm max}$, and an $H$-chord closing across a
matter chord contributes $q^{\Delta}$ while the crossing of the two matter
chords contributes $q^{\Delta_V\Delta_W}$. Because of the backward
Euclidean segments in (\ref{eq:Ffold0}) the intermediate vectors are
exponentially large compared to the final answer (their norms
grow like $e^{O(1/\lambda)}$),
so double precision arithmetic is insufficient;
we used multiple-precision arithmetic with dps decimal digits. We increased $n_{\rm max}$ and dps to achieve both convergence
and the desired precision of the extracted results. The largest values
used for the results shown here were $n_{\rm max}$ = dps = 440;
the required values scale approximately like $1/\lambda$.
We have checked that, once we take $n_{\rm max}$ large enough
for the chord sums to converge, the numerics are extremely stable, and
truncating the chord sum at that point has no effect within the precision of our analysis; this justifies calling our numerics ``exact DSSYK results''.

Typical results are shown in Figure~\ref{fig:OTOC975} for $q = 0.975$.
The standard OTOC $F_{\rm std}(t)$ has its maximum value at $t=0$,
and decreases monotonically thereafter. The real part of the folded OTOC
$F_{\rm fold}(t)$ has its minimum value at $t=0$, and
rises monotonically before peaking shortly before the scrambling time
(the modulus $|F_{\rm fold}(t)|$ peaks at $t_*$ itself, cf.\ the
remarks below (\ref{eq:tscram})); this is the anti-scrambling behavior.

While the standard OTOC
$F_{\rm std}(t)$ is strictly real, the folded OTOC $F_{\rm fold}(t)$
has a rapidly growing positive imaginary part near the scrambling time; this
is the exponentially small term $\pi w^2(w-1)e^{-w}$ of (\ref{eq:foldPhi}),
which becomes $O(1)$ at $a\sim 1/4$; it is consistent with
the expected behavior in
the mild backreaction regime where the de Sitter eikonal shockwave analysis
is also valid.


\begin{figure}[htb]
\begin{center}
\includegraphics[scale=.63]{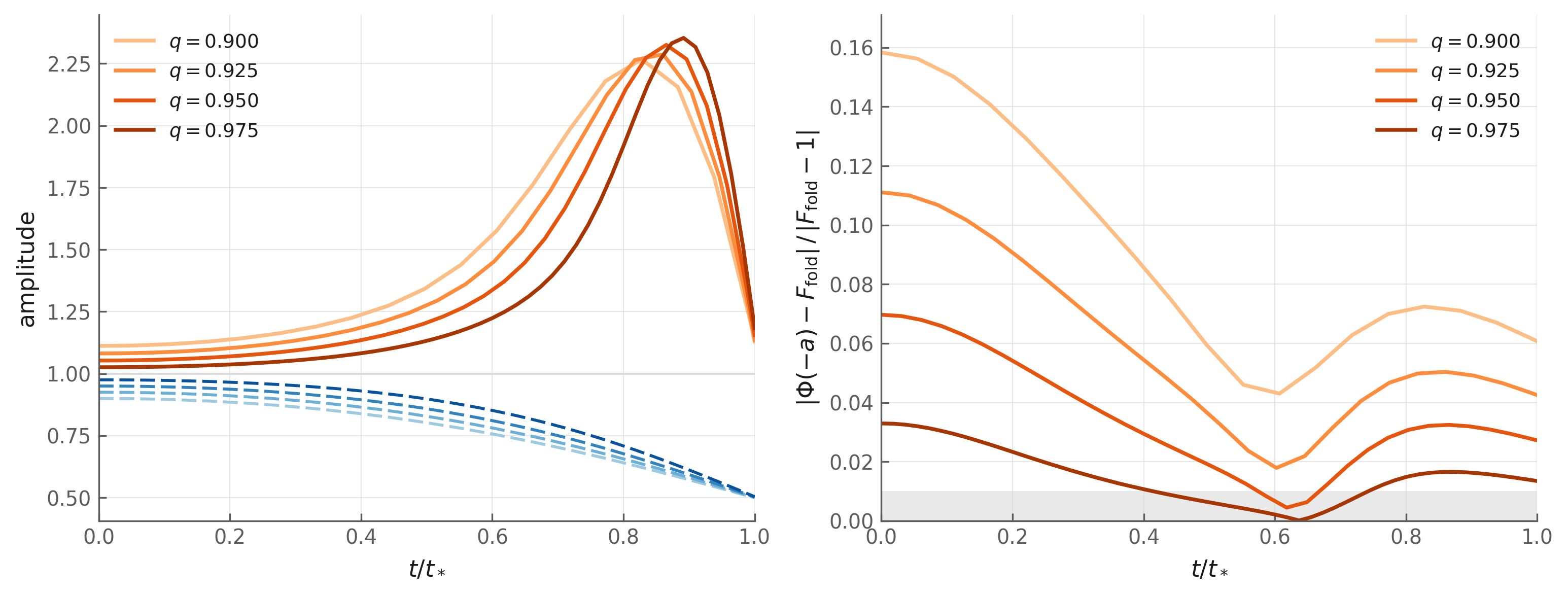}
\caption{{\bf Left:} $\lambda$ dependence of the real part of 
the folded OTOC $F_{\rm fold}(t)$ and the 
standard OTOC $F_{\rm std}(t)$,
in the exact DSSYK model,
for $q =$ 0.9, 0.925, 0.95, and 0.975, implying 
$\lambda =$ 0.1054, 0.078, 0.0513, and 0.0253. 
{\bf Right:} dependence on $\lambda$ of the magnitude of the fractional difference between the eikonal result $\Phi(-a(t))-1$ 
and $F_{\rm fold}(t)-1$,
where $a(t)$ was computed by matching $\Phi(a(t))$ to
$F_{\rm std}(t)$.}
\label{fig:OTOC_universal}
\end{center}
\end{figure}


The folded OTOC also has
a very small negative imaginary part at early times;
this has no counterpart in the eikonal
formula, and is a combination of two effects that we will now describe.

The first effect is that the 
numerical results for $F_{\rm fold}(t)$ are sensitive to
whether we perform the folding using the exact value of
$\beta_f = \pi/\Jf$, or an approximation to it.
In the language of Section~\ref{sec:rotation}, the results are sensitive to
any over-rotation of $z$ beyond $\pi$.
With this criterion we can define an ``exact'' $\beta_f$ by
requiring $F_{\rm fold}(0)$ to be stationary with respect to $\betaf$:
$\partial_{\beta_f}F_{\rm fold}(t)\vert_{t=0}=0$. To see what this means, 
let $f(\tau)\equiv\Tr\big(V(\tau)WV(\tau)W\big)$, which is real
for real $\tau$ and even in $\tau$ (Section~\ref{sec:fold}), so that (\ref{eq:Ffold0}) reads
\begin{equation}\label{eq:Ffoldf}
F_{\rm fold}(t) = f\big(\tfrac{\betaf}{2}+it\big),
\end{equation}
since $\betaf$ and $t$ enter only through the combination $\betaf/2+it$, and
\begin{equation}\label{eq:dbetaIm}
\frac{\partial\,{\rm Im}\,F_{\rm fold}}{\partial\betaf}
= -\frac12\,\frac{d\,{\rm Re}\,F_{\rm fold}}{dt}.
\end{equation}
Expanding about $t=0$,
${\rm Im}\,F_{\rm fold}(t) = f'(\betaf/2)\,t - f'''(\betaf/2)\,t^3/6+\dots$,
so the condition we used to define the exact $\beta_f$,
$\partial_{\betaf}F_{\rm fold}\vert_{t=0}=\tfrac12 f'(\betaf/2)=0$,
is precisely the condition that ${\rm Im}\,F_{\rm fold}$ has no term
linear in $t$. 

The same condition fixes the sign of the early growth. Since
\begin{equation}\label{eq:kin}
{\rm Re}\,F_{\rm fold}(t)=f(\betaf/2)-\tfrac12 f''(\betaf/2)\,t^2+O(t^4) ,
\end{equation}
and observing that semiclassically $f(\tau)=\Phi(a_\infty\cos\lambda_L\tau)$ and
$\tau=\betaf/2$ is the maximum of $f$ on the real axis, we see that
$f''(\betaf/2)<0$ and so ${\rm Re}\,F_{\rm fold}$ 
automatically rises from its value at $t=0$.
The initial rise of the
folded correlator thus follows from the definition of the fold depth as a local maximum of $f$ on the real axis; it is kinematic, and
would hold in any model in which such a maximum exists.
The dynamical content
of anti-scrambling is not this initial rise but rather
the details of the subsequent behavior: the exponent
$d\log a/dt\to2\pi/\betaf$, the shape $\Phi(-a(t)-i0)$ with the same
$a(t)$ as $F_{\rm std}$, and the sign of the $O(1)$ imaginary part,
all quantities that are compared below.

An error $\delta$ in $\betaf$ restores a linear term in 
${\rm Im}\,F_{\rm fold}$; from
(\ref{eq:dbetaIm}) and ${\rm Re}\,F_{\rm fold}\simeq1+4a+18a^2$ with
$a=\frac{\lambda}{4}\cosh(\lambda_Lt)$,
\begin{equation}\label{eq:dImdelta}
\delta\,{\rm Im}\,F_{\rm fold}\simeq
-\lambda^{3/2}\,\delta\,\sinh(\lambda_Lt)\,\big(1+9a\big),
\end{equation}
which, from the analysis in Section~\ref{sec:rotation},
is what we expect from an over-rotation of $z$ by $\lambda_L\delta/2$
beyond $\pi$, carried to second order in
the asymptotic series for $\Phi$.

It is useful to compare this exact $\beta_f$ to
(\ref{eq:Jexpand}), where we already have an expression
for $\beta_f$ accurate at next-to-leading order (NLO) in small
$\lambda$:
\begin{equation}
\beta_f = \frac{\pi}{\sqrt{\lambda}(1 + \lambda/8 + O(\lambda^2))}
\end{equation}
This agrees with the exact $\beta_f$
computed numerically as an expansion in $\lambda$:
\begin{align}
\beta_f = \frac{\pi}{\sqrt{\lambda}(1 + \lambda/8 + c_2\lambda^2 + c_3\lambda^3 + O(\lambda^4))}
\end{align}
where we find:
\begin{equation}
c_2 = -0.09635 \pm 0.00002\; ;\qquad
c_3 = 0.075 \pm 0.002
\end{equation}
For the plots and tables shown here, we have defined the fold using
the NNLO correction $c_2\lambda^2$ for $\beta_f$ but not including 
$c_3\lambda^3$; this is very
close to the exact fold, and its small residual effect is discussed below.

Equation (\ref{eq:dbetaIm}) also quantifies how much the comparison with the
eikonal amplitude depends on the choice of $\betaf$: a shift $\delta$ of
$\betaf$ changes the full correlator by 
$\delta F_{\rm fold}=-\tfrac{i\delta}{2}\,\partial_tF_{\rm fold}$. 
For example at the
scrambling time for $q=0.975$, changing $\betaf$ by its NLO term ($\delta\simeq0.06$, i.e.\ $0.3\%$)
shifts $|F_{\rm fold}|$ by about $1.5\%$. 
The agreement with $\Phi(-a-i0)$ at
the scrambling time is therefore uncertain at the percent level from the
definition of the fake period alone; the stationary $\betaf$ is the choice that is
defined without reference to the eikonal formula.

The second effect arises because,
even at the exact $\beta_f$, ${\rm Im}\,F_{\rm fold}$ has a small
negative imaginary part at early times.
As we already observed in Section~\ref{sec:fake},
at finite $\lambda$ the exact $\beta=0$ DSSYK correlators are only
approximately periodic in Euclidean time with period $\betaf$
(cf.\ the two-point function in (\ref{eq:betaf})), so $f(\tau)$ is not
exactly symmetric about $\tau=\betaf/2$ and $f'''(\betaf/2)\neq0$.
The residual ${\rm Im}\,F_{\rm fold}-{\rm Im}\,\Phi(-a-i0)$ therefore
starts at order $t^3$. 

Numerically we have modeled 
this early time non-eikonal behavior for $q=0.9$--$0.975$,
using the exact values of $\beta_f$.
We find:
\begin{equation}\label{eq:Imdip}
{\rm Im}\,F_{\rm fold}(t)-{\rm Im}\,\Phi(-a-i0)\simeq
-\frac{\lambda^3}{8}\,\sinh(\lambda_Lt)\big(\cosh(\lambda_Lt)-1\big)
\;\simeq\; -\frac{\lambda^{9/2}}{2}\,t^3 \quad(\lambda_Lt\ll1),
\end{equation}
i.e.\ the $\sinh\cosh$ structure of the second-order term, with its
$\sinh$ part fixed by the absence of a linear term. The $t^3$ coefficient
divided by $-\lambda^{9/2}/2$ is $1.33$, $1.16$, $1.08$, $1.05$ for
$q=0.9$, $0.925$, $0.95$, $0.975$, approaching one as $\lambda\to0$;
see the right plot of Figure~\ref{fig:OTOC_artifact}.
The functional form in (\ref{eq:Imdip}) holds to within $\sim\!5\%$ for
$\lambda_Lt\lesssim1.5$ ($t\lesssim0.41\,t_*$) at $q=0.95$. 
The dip in the left plot of Figure~\ref{fig:OTOC_artifact} is this term, cut off by the
onset of the physical ${\rm Im}\,\Phi(-a-i0)$; the NNLO $\betaf$ used
in the plots adds a small linear piece (\ref{eq:dImdelta}) that
deepens the dip by about 20\% at $q=0.925$. Like the rest of the
finite-$\lambda$ corrections, it vanishes as $\lambda\to0$.


To see that the anti-scrambling of the folded OTOC
$F_{\rm fold}(t)$ is consistent with de Sitter shock wave behavior,
we have performed the following exercise:
\begin{enumerate}
\item Assume $F_{\rm std}(t) = \Phi(a(t))$ and use the numerical results
to determine the monotonically increasing function $a(t)$.
\item Use equation (\ref{eq:Phi}) to compute $\Phi(-a(t))$
\item Compare Re$[\Phi(-a(t))]-1$ to Re$[F_{\rm fold}(t)]-1$,
and compare Im$[\Phi(-a(t))]$ to Im$[F_{\rm fold}(t)]$.
\end{enumerate}
This exercise tests the pure $a\to-a$ content of the fold, independently
of how accurately (\ref{eq:zcosh}) describes $a(t)$ at finite $\lambda$.

The results are shown in Tables~\ref{tab:q975}, \ref{tab:q950} and \ref{tab:q900} for three values of
$\lambda$. The real and imaginary part values agree well, and, as seen in the tables and
in Figure~\ref{fig:OTOC_universal},
the agreement improves as $\lambda$ decreases, as expected.
For the smallest $\lambda$ value shown, the agreement 
at the scrambling time of the magnitudes 
and of the imaginary parts is at the percent level;
the fractional agreement of the real parts is worse since the real parts are decreasing rapidly towards zero at the scrambling time.
The good agreement of the imaginary parts confirms the sheet selection of Section~\ref{sec:rotation}.

We also see in the tables that the extracted $a(t)$ increases exponentially
such that, as we approach the scrambling time,
$d\,$log$(a)/dt$ approaches the fake temperature Lyapunov
exponent chaos bound $2\pi/\betaf \simeq 2\sqrt{\lambda}$; the fold does
not change the exponent, in agreement with
the results
found in~\cite{harlow2026antiscramblingeuclideanfoldsobserver}
for de Sitter observer correlators.


\subsection{Short time behavior}\label{sec:short}
As $t \to 0$, $F_{\rm std}(t) \to q$; this is exact, because the
only remaining connected chord diagram is the two matter chords
crossing with weight $q^{\Delta_V\Delta_W} = q$.
Numerically we also find $F_{\rm fold}(0) \simeq q^{-1}$; this is not
exact, but holds through $O(\lambda^2)$, as the following matching shows.

Rewriting the long-time expression (\ref{eq:zcosh}) for $a(t)$
as $a(t) = a_\infty\,$cosh$(2\sqrt{\lambda}t)$ with
$a_\infty = \lambda/4$, we can match $F_{\rm std}(t)$ at
short times to the expansion $1-4a +18a^2+\ldots$, and
match $F_{\rm fold}(t)$ at
short times to the expansion $1+4a +18a^2+\ldots$, to get
\begin{align}
F_{\rm std}(t) &\simeq \Phi\big(a_{\rm std}\,{\rm cosh}(2\sqrt{\lambda}t)\big)
\; ;\qquad a_{\rm std} = a_\infty (1 + \frac{5}{8}\lambda)
\nonumber\\
F_{\rm fold}(t) &\simeq \Phi\big(-a_{\rm fold}\,{\rm cosh}(2\sqrt{\lambda}t)\big)
\; ;\qquad a_{\rm fold} = a_\infty (1 - \frac{5}{8}\lambda)
\end{align}
where indeed $\Phi(a_{\rm std}) = 1-\lambda+\lambda^2/2+O(\lambda^3) = q + O(\lambda^3)$
and $\Phi(-a_{\rm fold}) = 1+\lambda+\lambda^2/2+O(\lambda^3) = q^{-1}+O(\lambda^3)$.
Thus at short times it is not true that $F_{\rm std}$ and
$F_{\rm fold}$ are simply related by $a(t) \to -a(t)$, but the
discrepancy is at next-to-leading order in $\lambda$, and
it shrinks towards the scrambling regime as seen in
Tables~\ref{tab:q975}--\ref{tab:q900}.



\begin{figure}[htb]
\begin{center}
\includegraphics[scale=0.71]{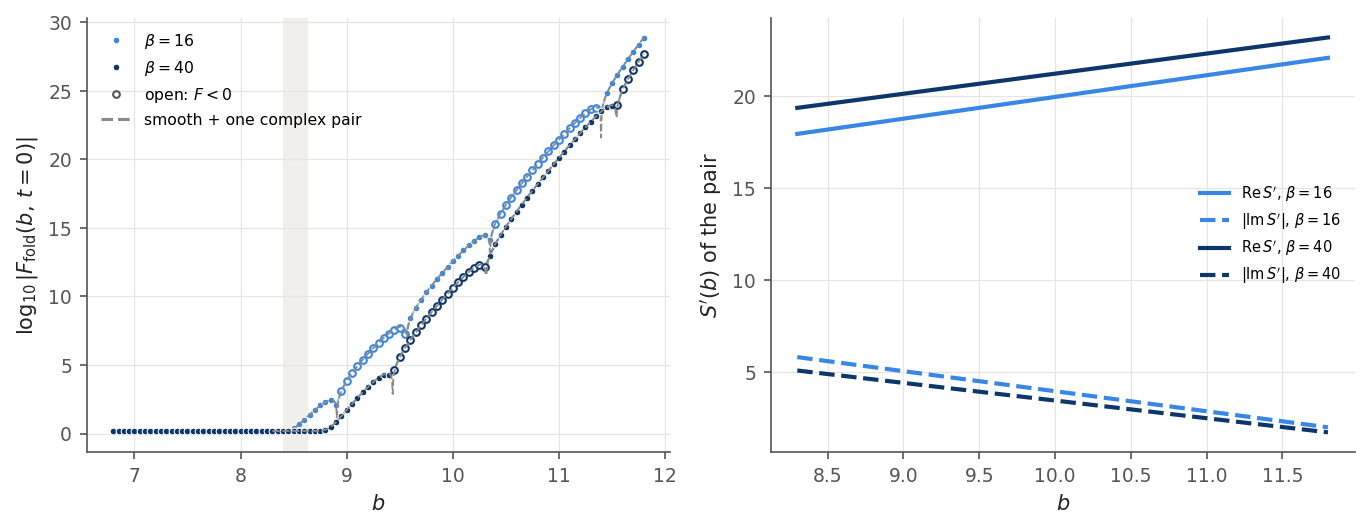}
\caption{The blowup for $q$$=$$0.925$, $\lambda$$=$$0.078$. {\bf Left:} Log of the
DSSYK folded OTOC $|F_{\rm fold}(t$$=$$0)|$ as a function of
the effective backwards fold $b = s_d$$-$$\beta/4$, where the fold $s_d$ is taken
as a free parameter.
The numerics are computed up to $n_{\rm max}$$=$$340$ with dps$=$$200$ and converge for all points shown.
We see that $|F_{\rm fold}(t$$=$$0)|$ is $O(1)$ for values of $b < b_c$,
where $b_c \simeq 8.7 \simeq 1.6\beta_f/2$,
and then blows up exponentially, with an oscillation, for larger folds. The dashed line
shows a fit to the form
$F_{\rm fold}(t$$=$$0) = F_{\rm smooth}(b) + \e^{c(b)}\,{\rm cos}\,\theta(b)$
where, for each value of $\beta$, 
$F_{\rm smooth}(b)$, $c(b)$, and $\theta(b)$ are fit to quadratic functions of $b$.
{\bf Right:} Assuming that the blowup corresponds to a complex conjugate pair
of saddles, 
$F_{\rm fold}(t$$=$$0) = F_{\rm smooth}(b)$$+$$\e^{S(b)}$$+$$\e^{S^*(b)}$,
we extract Re$S^\prime(b)$ and $|$Im$S^\prime(b)|$.
}
\label{fig:fb_saddles}
\end{center}
\end{figure}


\begin{figure}[htb]
\begin{center}
\includegraphics[scale=0.71]{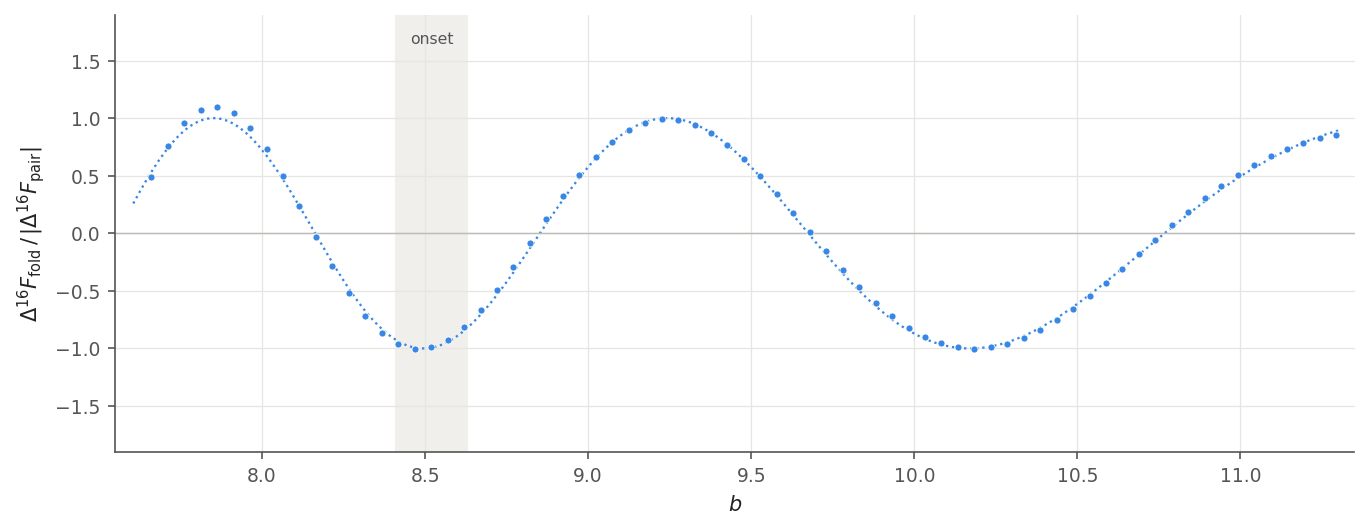}
\caption{Anti-Stokes behavior for $q$$=$$0.925$, $\lambda$$=$$0.078$, $\beta$$=$$16$,
taking the backward Euclidean segment $b=s_d$$-$$\beta/4$ as a free parameter.
The data are the exact $F_{\rm fold}$ at $t$$=$$0$ on a uniform grid with spacing $h=0.05$. 
Above the onset (shaded band, where the
exponentially growing term rises from 1\% of the $O(1)$ correlator to equal
it) the correlator is dominated by a complex-conjugate pair of saddles,
$F_{\rm fold}(t$$=$$0)\simeq{\rm Re}\,F_{\rm pair}$,
$F_{\rm pair}(b)=2\,e^{S(b)}$,
with $S(b)$ quadratic in $b$
and fitted for $b\ge8.3$.
The complex saddle pair in the folded OTOC continues smoothly
below its onset.
Points: $\Delta^{16}F_{\rm fold}/|\Delta^{16}F_{\rm pair}|$, where
$\Delta^{16}$ is the 16th finite difference on the $b$ grid. 
This reduces
the slowly varying part of $F_{\rm fold}$ to a residual floor of
$\sim7\times10^{-10}$ (for $|F_{\rm fold}|\simeq1.26$), while the pair, which grows
by $e^{h\,{\rm Re}\,S'}\approx2.4$ per step, survives.
Each point is placed where the pair's contribution peaks.
Dotted: $\cos\arg\Delta^{16}F_{\rm pair}$, i.e.\ the pair extrapolated from
above the onset with unit coefficient.
Below the onset the points follow the same oscillation down to $b\simeq7.7$,
where the pair is $\sim4\times10^{-7}$ of the smooth part: a two-parameter fit
$\Delta^{16}F_{\rm fold}=A\,{\rm Re}\,\Delta^{16}F_{\rm pair}+B\,{\rm Im}\,\Delta^{16}F_{\rm pair}$
on $7.66\le b\le8.37$ gives $A=1.02$ and $B=-0.04$, compared with
$A=1$, $B=0$ above the onset by construction.
}
\label{fig:fb_wiggles}
\end{center}
\end{figure}


\begin{figure}[htb]
\begin{center}
\includegraphics[scale=0.82]{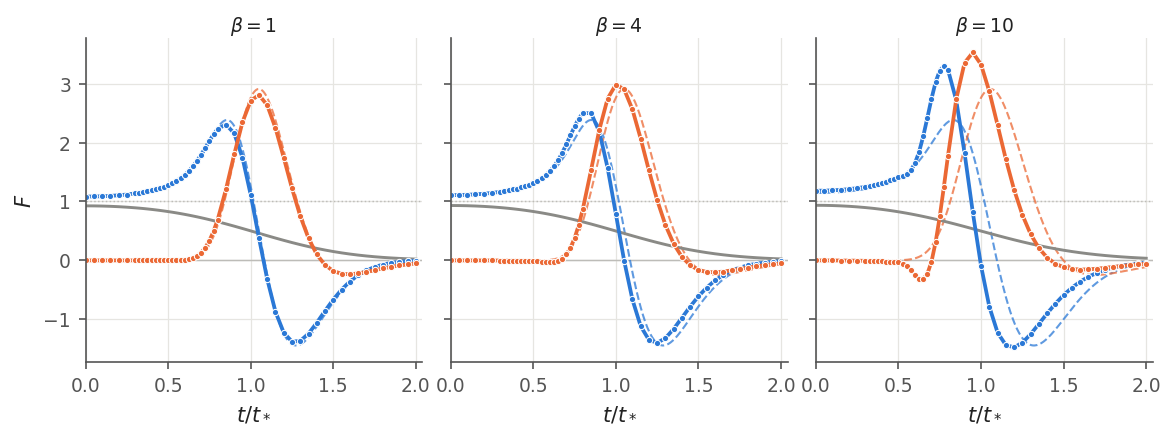}
\caption{High temperature examples of
the folded DSSYK OTOC $F_{\rm fold}(t)$
for $q$$=$$0.925$, $\lambda$$=$$0.078$.
For each plot we use the finite $\beta$ folding
prescription, with the fold $s_d$ determined
by requiring $\partial_{s_d} F_{\rm fold}(t)\vert_{t=0}=0$.
For these examples with $\beta = 1,$ 4, 10, we set
$s_d = 5.62,$ 6.24, 8.49; by comparison the leading order
in small $\lambda$ formula $s_d = \beta/(2v)$ gives
$s_d = 5.68,$ 6.39, 8.88. The real and imaginary parts
of $F_{\rm fold}(t)$ are plotted separately (solid blue and orange lines); also shown
is the unfolded OTOC $F_{\rm std}(t)$ (gray line), and the real and imaginary parts of the eikonal function $\Phi(-a(t)-i0)$ (dashed blue and orange lines) where $a(t)$ was computed by matching $\Phi(a(t))$ to
$F_{\rm std}(t)$.}
\label{fig:finitebeta}
\end{center}
\end{figure}


\begin{figure}[htb]
\begin{center}
\includegraphics[scale=0.73]{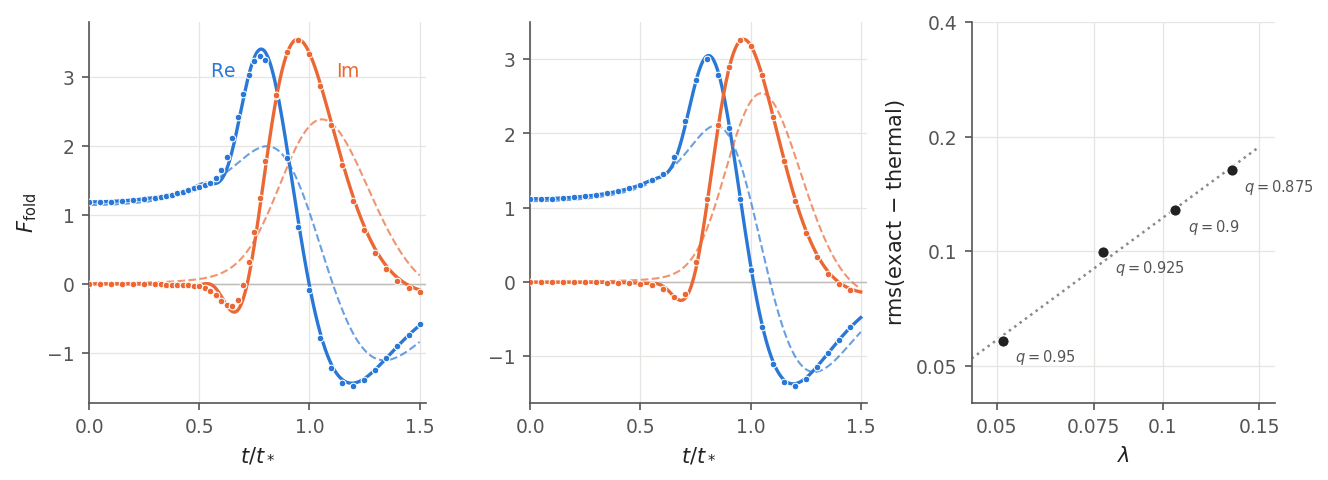}
\caption{The folded OTOC at finite temperature, for fixed
$\beta\sqrt{\lambda}=2.79$, compared with the long-time
limit of~\cite{mukhametzhanov2023largepsykchord}.
{\bf Left:} $q=0.925$, $\beta=10$ ($\lambda=0.078$), $s_d=8.49$; {\bf Middle:} $q=0.95$,
$\beta=12.33$ ($\lambda=0.051$), $s_d=10.63$. 
Points: the exact ${\rm Re}\,F_{\rm fold}$ (blue) and ${\rm Im}\,F_{\rm fold}$
(orange). Dashed: the eikonal-like result of (\ref{eq:Mz}) with the average
energy fixed to its saddle point.
Solid: the same expression averaged over the thermal distribution of the
energy parameter $s=\theta/\lambda$ across which the Lyapunov exponent
$\lambda_L$ varies,
using $\lambda_L(\theta ) = 2\sqrt\lambda\,\mathrm{sin}\,\theta$.
{\bf Right:} the remaining deviation
${\rm rms}(\text{exact}-\text{thermal})$,
with the rms taken over $0\le t\le1.5\,t_*$, as a function of $\lambda$ at
fixed $\beta\sqrt\lambda=2.79$, for $q=0.875$, $0.9$, $0.925$ and $0.95$. The dotted line is a power law fit to the four points.}
\label{fig:finitebeta_thermal}
\end{center}
\end{figure}


\subsection{Anti-Stokes behavior of the thermal fold at high temperatures}
\label{sec:stokes}
In Section~\ref{sec:fail} we saw from the integral defining the DSSYK OTOC that
it should be possible to construct a folded OTOC with approximately de Sitter-like
eikonal behavior for some range of finite $\beta$, by using a fold of approximately $\beta/(2v)$
instead of $\beta_f/2$. However, we also saw from the saddle analysis that we expect
a problem for folds exceeding roughly $1.5\,\beta_f/2$. Here we will use numerical results
to quantify this and identify the source of the breakdown for too-large folds.

The left plot in Figure~\ref{fig:fb_saddles} shows $|F_{\rm fold}(t$$=$$0)|$ as a function of
the effective backwards fold $b = s_d$$-$$\beta/4$, where the fold $s_d$ is taken
as a free parameter. In this example $q$$=$$0.925$ corresponding to $\lambda$$=$$0.078$,
and results are shown for $\beta$$=16$ and 40.
$|F_{\rm fold}(0)|$ is $O(1)$ for values of $b < b_c$, 
where $b_c \simeq 8.7 \simeq 1.6\,\beta_f/2$,
and then blows up exponentially, with an oscillation, for larger folds. The dashed line
shows a fit to the form
$F_{\rm fold}(t$$=$$0) = F_{\rm smooth}(b) + \e^{c(b)}\,{\rm cos}\,\theta(b)$.
Assuming that the blowup corresponds to a complex conjugate pair
of saddles, 
$F_{\rm fold}(t$$=$$0) = F_{\rm smooth}(b)$$+$$\e^{S(b)}$$+$$\e^{S^*(b)}$,
we extract Re$\,S^\prime(b)$ and $|$Im$\,S^\prime(b)|$ as shown in the right plot.
We see that Re$\,S^\prime(b)$ is slowly increasing with $b$ and is only weakly dependent on the value
of $\beta$. Since the integrand of (\ref{eq:exactOTOC}) depends on the fold depth
through e$^{s_d(E_2+E_4-E_1-E_3)}$, Re$\,S^\prime(b)$ is the total energy splitting between
the backward and forward faces at the saddle; it is heading towards its maximal
value $4E_0$ with the two pairs of faces at opposite edges of the spectrum.
The decrease of $|$Im$\,S^\prime(b)|$ with larger $b$
indicates that the conjugate pair of saddles approach the real axis as $b$ increases.

For the $\beta$$=$$16$ example, if we compute the fold $s_d$ from $\beta/(2v)$,
or if we determine it by requiring $\partial_{s_d} F_{\rm fold}(t)\vert_{t=0}=0$,
we get $b = 7.7$ or 6.9, respectively; thus with our finite $\beta$ folding
prescription the $\beta$$=$$16$ folded OTOC, and by extension the folded OTOC for any smaller value of $\beta$, has $b<b_c$. For the $\beta$$=$$40$ folded OTOC, with the fold defined as $\beta/(2v)$,
we get $b = 13.6$, well above $b_c$; for this too-large $\beta$ the eikonal behavior cannot
be retrieved from the folded OTOC. With the fold $\beta/(2v)$,
$b$ reaches $b_c$ at $\beta\simeq20$, i.e.\ $\beta\sqrt\lambda\simeq5.6$.

Figure~\ref{fig:fb_wiggles} shows in more detail what is happening with the complex saddles.
Here we have plotted for the $\beta$$=$$16$ example just the oscillating part $F_{\rm pair}(b)$
of the folded OTOC $F_{\rm fold}(t$$=$$0)$, again taking the fold as a free parameter. The oscillating part is extracted
using the finite difference method.
Here we see that the oscillatory behavior is also present for $b$$<$$b_c$, over many orders of magnitude
of $|F_{\rm pair}(b)|$, and with approximately the same coefficient and phase as we got by fitting
the behavior for $b$$>$$b_c$. This indicates that the blowup is anti-Stokes behavior~\cite{bender1999advanced}:
the coefficients of the saddles are approximately constant on both sides of the crossover
line $b=b_c$, so the blowup is the pair of saddles 
becoming dominant over
the original symmetric saddle discussed in Section~\ref{sec:longtime}.

In Figure~\ref{fig:finitebeta} we construct
the folded OTOC $F_{\rm fold}(t)$ with the fold determined by requiring $\partial_{s_d} F_{\rm fold}(t)\vert_{t=0}=0$,
and plot it versus time for three $q=0.925$ examples: $\beta =1$, 4, and 10. 
Here we have defined the scrambling time as 
$t_* = \mathrm{arccosh}(\lambda_L / 2\lambda^{3/2}) / \lambda_L$,  with $\lambda_L = 2\pi v/\beta$,
consistent to leading order with~\cite{harlow2026antiscramblingeuclideanfoldsobserver}
and with (\ref{eq:tscram}) for $\beta\to 0$.
For these
examples $b$ is well below $b_c$$=$$8.7$, e.g.\ for $\beta$$=$$10$ we have $b$$=$$6.0$.
For all three examples a numerical estimate of the contribution from the complex saddles predicts them to be
exponentially suppressed over the entire time range $0$$<$$t$$<$$2t_*$, and indeed no sign of them is
visible in the plots of the exact points.
The first example $\beta$$=$$1$ tracks the expected de Sitter-like eikonal behavior
out to $t=2t_*$ at about the same level of precision as our previous $\beta$$=$$0$
results. However the same plot for $\beta$$=$$10$ (and to a lesser extent for $\beta$$=$$4$) 
shows substantial deviations.

Figure~\ref{fig:finitebeta_thermal} shows the origins of the deviations
in the $\beta$$=$$10$ example. Here we fix $\beta\sqrt{\lambda}=2.79$,
corresponding to the $\beta$$=$$10$, $\lambda$$=$$0.078$ case of Figure~\ref{fig:finitebeta},
and compute the exact folded OTOC for four different pairs of $\beta$,$\lambda$ values,
all of them corresponding to $b$$<$$b_c$.
We can then compare these with the folded version of the long-time
limit of DSSYK, which gave the eikonal-like result (\ref{eq:Mz}).
Here we observe that eq.~(5.20) of~\cite{mukhametzhanov2023largepsykchord}
was derived making the approximation of fixing the average
energy $s$$=$$(s_1+s_2+s_3+s_4)/4$
to its saddle point value. At finite temperature this energy 
$s$$=$$\theta/\lambda$ has a thermal
distribution that affects the Lyapunov exponent, 
i.e.\ $\lambda_L(\theta )= \lambda\,E_0\,\mathrm{sin}\,\theta \to 2\sqrt\lambda\,\mathrm{sin}\,\theta$ at leading
order,
which in turn affects the fold rotation of $z$ in 
(\ref{eq:zrot}) which is $\lambda_L(\theta)s_d$
with $\lambda_L(\theta_s)s_d = \pi$ at the saddle
value $\theta_s$. Thus for
a fixed fold $s_d$, we expect a thermal
spread in the fold rotation angle of order the
saddle standard deviation $\sigma_\theta$ from (\ref{eq:variance})
times
$d\lambda_L/d\theta|_{\theta=\theta_s}\,s_d$,
which is $\pi\,\mathrm{cot}\,\theta_s\,\sigma_\theta$, of order $\sqrt\lambda$.

The solid lines
in the left and middle panels of Figure~\ref{fig:finitebeta_thermal} were obtained by performing the thermal average over $\rho_q(s)\,\e^{-\beta E(s)}$ on a grid $\pm 8$ sigma around the saddle, weighted by $\sin^4(\lambda s)$ and normalized by $\langle\sin^2(\lambda s)\rangle^2$ from the two-point functions,
to account for the prefactors in eqs.~(5.17) and (4.12) of~\cite{mukhametzhanov2023largepsykchord}.
For the part of the thermal distribution with $\lambda_L(\theta)\,s_d > \pi$, we
continued analytically through the cut on the negative real axis as required by analyticity in the fold depth at fixed energy.
Clearly the solid lines agree much better with the exact points.
The small remaining deviation of the exact points from the thermally averaged form falls roughly as $\lambda^{1.1}$ at fixed $x$ (right panel of Figure Figure~\ref{fig:finitebeta_thermal}), consistent with additional finite $\lambda$ effects.

We conclude that, taking into account the thermal spread
in the fold
rotation angle, the $\beta$$=$$0$ folding prescription successfully extends to finite
$\beta$ for sufficiently high temperatures such that we satisfy the condition
$b<b_c\simeq1.6\,\betaf/2$ on the fold depth. 
Above $b_c$ the conjugate saddle pair
dominates already at $t=0$, which is
the anti-Stokes crossover of Figures~\ref{fig:fb_saddles}
and~\ref{fig:fb_wiggles}. 

\section{Late time behavior and the two-shell dilaton
invariant}

Harlow and Zhao derive the eikonal formula from the time advance
of a probe crossing a shell, linearized in the shell energy
(their eq.~(3.34)); the neglected terms become $O(1)$ at twice the
scrambling time, thus their four-point function is stated to hold for
``milder" (in terms of backreaction) anti-scrambling in the regime
$1\ll|t_1-t_2|<2t_*$ (their eq.~(4.29)). 

\begin{figure}[htb]
\begin{center}
\includegraphics[scale=.95]{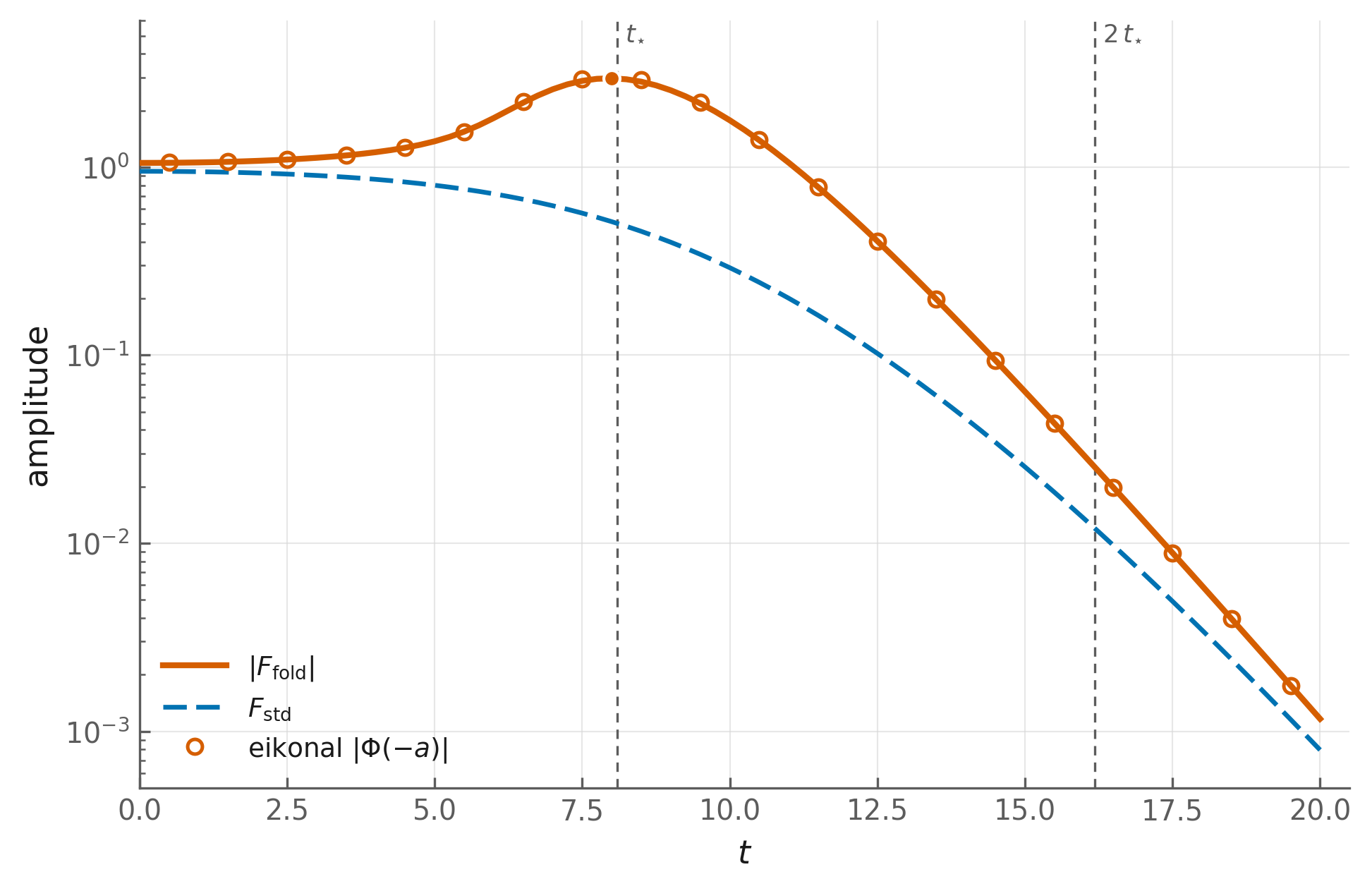}
\caption{Late-time behavior of the folded OTOC. Plotted are $|F_{\rm fold}|$ (solid line),
$F_{\rm std}$ (dashed line), and
eikonal $|\Phi(-a)|$ (circles), with $a(t)$ extracted from
$F_{\rm std}$, versus time
for $q=0.95$, corresponding to $\lambda=0.0513$, $\beta_f=13.786$, and scrambling time $t_*=8.09$.}
\label{fig:late}
\end{center}
\end{figure}

\begin{figure}[htb]
\begin{center}
\includegraphics[scale=.77]{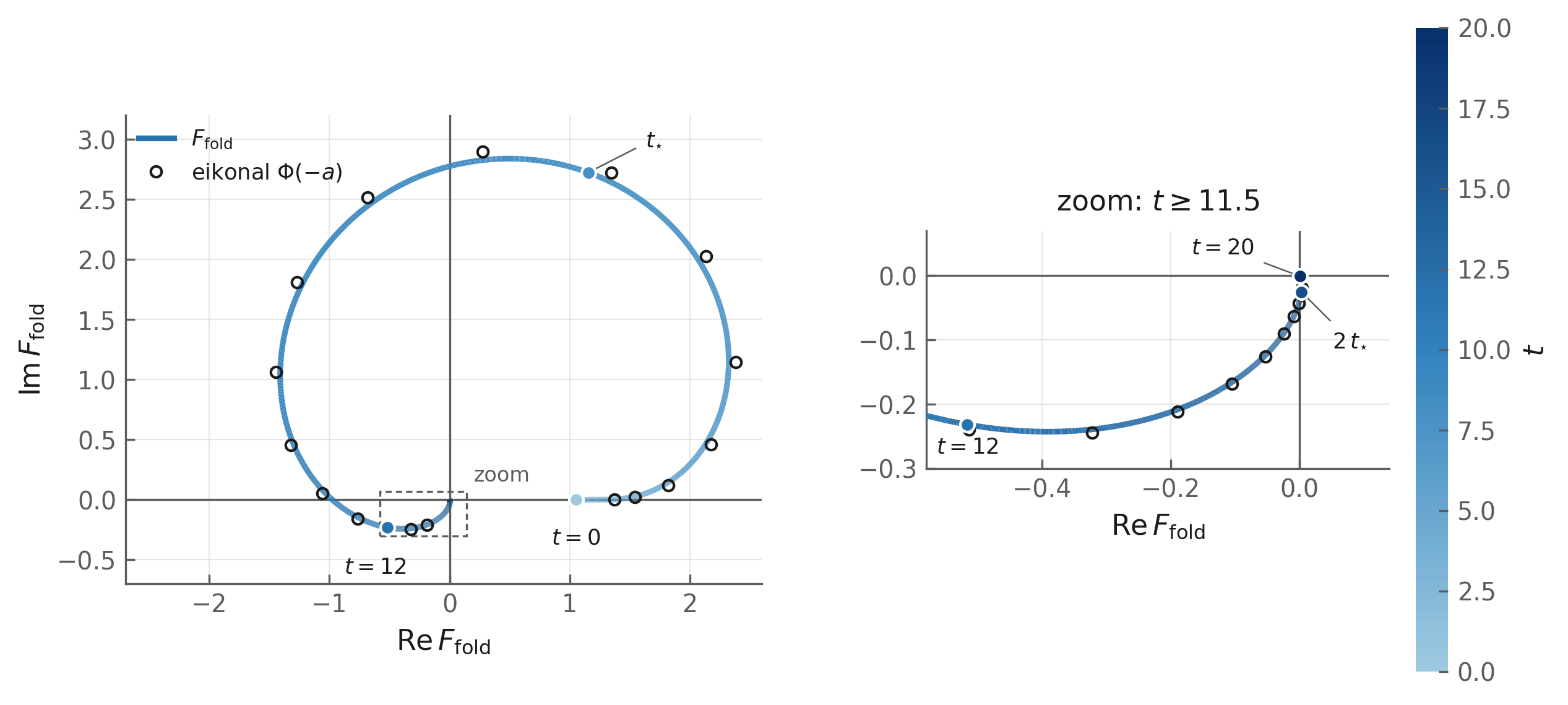}
\caption{Late-time behavior of the folded OTOC in the complex plane.
Plotted are the trajectory of $F_{\rm fold}(t)$ (solid line, colored by $t$)
and the
eikonal $\Phi(-a(t)-i0)$ (circles), with $a(t)$ extracted from
$F_{\rm std}$, 
for $q=0.95$, corresponding to $\lambda=0.0513$, $\beta_f=13.786$, and scrambling time $t_*=8.09$. The inset zooms on $t\ge11.5$, where ${\rm Im}\,F_{\rm fold}$ has turned negative and approaches zero as $-\pi/a^2$.
}
\label{fig:latephase}
\end{center}
\end{figure}

The late time behavior of $F_{\rm fold}(t)$ 
is shown in
Figures~\ref{fig:late} and \ref{fig:latephase}.
As before we extract $a(t)$ from $F_{\rm std}(t)$ to compute $\Phi(-a(t))$. We find that
the extracted $a(t)$ remains exponential:
$d\log a/dt$ settles to within $0.4\%$ of $2\pi/\betaf$,
with no change at $2t_*$.
We see that $F_{\rm fold}(t)$ continues to track $\Phi(-a(t))$
at late times; its magnitude decreases monotonically after
peaking at $t_*$ and it has the same asymptotic behavior 
as $F_{\rm std}(t)$.

Harlow and Zhao in Section 3
of~\cite{harlow2026antiscramblingeuclideanfoldsobserver} 
mapped $2+1$-dimensional gravity with $\Lambda>0$ and spherical dust shells onto JT gravity
with positive cosmological constant, and solved 
the problem of two colliding shells at the classical level. 
They expressed their solution in terms of $\kappa_{\rm JT}$, the JT gravity version of the dilaton invariant defined for
sine-dilaton gravity in (\ref{eq:kappasd}).
$\kappa_{\rm JT}$ takes different values in each of the four regions cut out by the shells, and in one of the
post-collision regions their solution contained a term growing 
exponentially with the time separation $t_2$ of the two shells,
\begin{equation}
\kappa_{\rm JT} \simeq \kappa_{\rm JT}^0 + 2\sqrt{\kappa_{\rm JT}^0}\,\omega_2 - 2\omega_1\omega_2\cosh t_2,
\label{eq:HZgrow}
\end{equation}
their eq.~(3.45) expanded to second order
in $\omega_i/\sqrt{\kappa_{\rm JT}^0}$, where
$\kappa_{\rm JT}^0$ is the unperturbed value of the dilaton invariant.
The first-order term is the shift produced by a single shell, of
relative size $O(\omega/\sqrt{\kappa_{\rm JT}^0})$; the other regions
have only shifts of this kind. The bilinear term is the one that grows:
at $-t_2=2t_{\rm scr}+O(1)$, with $t_{\rm scr}=\log S_{\rm dS}$ (their
eqs.~(3.33) and (3.46)), it overwhelms $\kappa_{\rm JT}^0$, $\kappa_{\rm JT}$
changes sign, the dilaton becomes timelike, and this
post-collision region ends in a big crunch.
The growing term is the same
gravitational time advance that produces anti-scrambling in the
DSSYK folded OTOC; the crunch is
its backreaction followed to the end.

Since DSSYK has a bounded spectrum, we expect that the two-shell computation should have a DSSYK counterpart where the
would-be space-time singularity is replaced with some other
kind of behavior. In this section we identify this
counterpart and study it for the case of 
both light and heavy probes.

We can extract the unperturbed dilaton invariant
$\kappa_{\rm JT}^0$ from the sine-dilaton discussion
of Section~\ref{sec:sinedg}. Assume a geodesic observer at
$r$$=$$\pi$ and expand the classical solution of the
sine-dilaton field $\Phi_{\rm sd}$ near the observer as $\Phi_{\rm sd} = \pi + \rho$.
Then, ignoring boundary terms, the sine-dilaton action (\ref{eq:sdaction})
reduces to:
\begin{align}\label{eq:sdaction2}
\frac{1}{2\lambda}\left[
\int\,dt\,dr\,
(\pi R + \rho R +2\,{\rm sin}(\pi + \rho)) \right]
\end{align}
Ignoring the topological term $\pi R$, this is the action of
two-dimensional JT gravity with a positive cosmological constant:
\begin{align}\label{eq:sdaction3}
\frac{1}{2}\left[
\int\,dt\,dr\,
\Phi_{\rm JT}(R - 2) \right]
\end{align}
where we have identified the JT dilaton as $\rho/\lambda$.
We can now compute the JT dilaton invariant:
\begin{align}\label{eq:kappaJT}
\kappa_{\rm JT}(\theta ) = (\nabla \Phi_{\rm JT})^2
+ \Phi_{\rm JT}^2 = \frac{F}{\lambda^2} + \frac{\rho^2}{\lambda^2}
\simeq \frac{2}{\lambda^2}(1 + {\rm cos}\,\theta)
= \frac{2}{\lambda^2}\left( 1 - \frac{E}{E_0} \right)
\end{align}
Although we computed $\kappa_{\rm JT}(\theta )$ in the region
$\rho \ll 1$ near the observer, it differs from the sine-dilaton invariant
$\kappa_{\rm sd}(\theta )$ only by constants, so in fact
(\ref{eq:kappaJT}) applies in the entire bulk region; it
is the dilaton invariant
of~\cite{harlow2026antiscramblingeuclideanfoldsobserver}.
Notice that the big crunch behavior corresponds to 
$\kappa_{\rm JT}$ decreasing to zero and becoming negative,
which requires the energy becoming larger 
than $E_0$~\cite{Blommaert_2025,okuyama2025sitterjtgravitydoublescaled}.

The two-shell analysis in DSSYK corresponds to
a correlator with four matter insertions and thus four gaps:
\begin{align}
F(g_1,g_2,g_3,g_4) = \Tr \left[
\e^{-g_1 H} \,V\, \e^{-g_2 H}\,W\, \e^{-g_3 H} \,V\,\e^{-g_4 H} \,W
\right]
\end{align}
From the exact DSSYK expression (\ref{eq:exactOTOC}) we see
that the chord expansion assigns an energy $E_n$  with Boltzmann
weight e$^{-g_nE_n}$ to each of the
four faces of the crossed chord diagram, with the expectation
values given by a spectral representation of the form:
\begin{align}\label{eq:expEn}
\langle E_n \rangle = -\frac{\partial }{\partial g_n}{\rm log}F
= \frac{\int_0^\pi d\theta\, E(\theta) I_n(\theta) }
{\int_0^\pi d\theta\, I_n(\theta)}
\end{align}
where $I_n(\theta )$ is the exact energy distribution,
which in general is complex.

For our DSSYK OTOC $F_{\rm fold}$ we take the gaps
$(g_1,g_2,g_3,g_4)$ =
$(-\beta_f/2-it,\,\beta_f/2+it,\,-\beta_f/2-it,\,\beta_f/2+it)$.
These are the gaps of Section~\ref{sec:longtime} with all signs
reversed, which gives the same correlator by the evenness of $f$
(Section~\ref{sec:fold}); with this labeling faces 1 and 3 are the
backward-evolved ones, with Boltzmann weight e$^{+(\betaf/2+it)E}$.
Thus:
\begin{align}\label{eq:efold}
\langle E_1 \rangle = \langle E_3 \rangle 
= -\langle E_2 \rangle = -\langle E_4 \rangle
\equiv \langle E_{\rm fold}\rangle(t)\, ,
\qquad 
\langle E_{\rm fold}\rangle(t) = -\frac{i}{4}
\frac{d}{dt}\,{\rm log}\,F_{\rm fold}(t)
\end{align}

In the regime $a\ll1$, where $F_{\rm fold}$ tracks
$\Phi(-a(t))$, restoring the dependence on the conformal dimensions
$\Delta_V$$=$$\Delta_W$$=$$\Delta$ and working to leading order in small
$\lambda$:
\begin{align}
{\rm log}\,\Phi(-a(t)) \simeq
\lambda\,\Delta^2\, {\rm cosh}(\lambda_L t)
\simeq \frac{1}{2} \lambda\,\Delta^2\, {\rm e}^{2\sqrt{\lambda} t}
\end{align}
Using (\ref{eq:efold}) and writing $\kappa(t)$
= $\kappa_0 + \Delta\kappa(t)$,
with $\kappa_0 = 2/\lambda^2$,
we can compare the DSSYK result to the two-shell result
of~\cite{harlow2026antiscramblingeuclideanfoldsobserver}:
\begin{equation}
\frac{\Delta\kappa}{\kappa_0}=-\frac{\langle E\rangle_{\rm fold}}{E_0}\simeq \frac{i\lambda^2\Delta^2}{8}\,e^{\lambda_L t}
\qquad\longleftrightarrow\qquad
\frac{\Delta\kappa_{\rm JT}}{\kappa_{\rm JT}^0}\simeq-\frac{\omega_1\omega_2}{\kappa^0_{\rm JT}}e^{|t_2|},
\label{eq:match}
\end{equation}

Matching terms we get $|t_2|=\lambda_L t=(2\pi/\betaf)\,t$ to leading
order in small $\lambda$, consistent with $S_{\rm dS}=2/\lambda$ and our
definition (\ref{eq:tscram}) of the scrambling time; the shell energies match as
$\omega_1\omega_2\leftrightarrow\lambda^2\Delta^2\kappa_0/8=\Delta^2/4$,
i.e.\ $\omega\simeq\Delta/2$, thermal-scale shells with
$\omega/\sqrt{\kappa_0}=O(\lambda\Delta)$, the regime
of~\cite{harlow2026antiscramblingeuclideanfoldsobserver}.
The putative crunch time in DSSYK time units is given by:
\begin{equation}
t_{\rm crunch}=2t_*+\frac1{\lambda_L}\log\frac{2}{\Delta^2}
\label{eq:tcl}
\end{equation}
which is the classical extrapolation of the scrambling-regime growth
and, e.g.\ for $q$$=$$0.95$, lies about 10\% past $2t_*$.

From this exercise we see that the DSSYK analysis
of $F_{\rm fold}$ matches up structurally with the classical two-shell
but with crucial differences.
The DSSYK quantity in (\ref{eq:match}) is imaginary, while the two-shell
$\Delta\kappa_{\rm JT}$ is real and negative. Writing
$F_{\rm fold}=|F_{\rm fold}|\,e^{i\varphi}$, (\ref{eq:efold}) gives
${\rm Re}\,\langle E\rangle_{\rm fold}=\tfrac14\,d\varphi/dt$ and
${\rm Im}\,\langle E\rangle_{\rm fold}=-\tfrac14\,d\log|F_{\rm fold}|/dt$:
the growing term comes from the growth of the modulus of the folded
correlator, and appears as an imaginary part, i.e.\ a phase, of $\kappa$.
The match is therefore of
the growth exponent and of the magnitude of the two-shell term. The real part of
$\kappa$, which in the classical analysis decreases through zero at
$2t_{\rm scr}$, here only dips to a positive minimum,
${\rm Re}\,\kappa/\kappa_0=1-O(\lambda\Delta)$. In the eikonal regime,
to leading order in small $\lambda$,
\begin{equation}
\min_t{\rm Re}\,\kappa/\kappa_0\simeq1-0.40\lambda
\label{eq:sat}
\end{equation}
for $\Delta=1$. With the identification of
$\langle E\rangle_{\rm fold}$ as the dilaton invariant, the folded
correlator for thermal-scale probes does not reproduce the two-shell
backreaction of~\cite{harlow2026antiscramblingeuclideanfoldsobserver};
the growing term is present as a phase of $\kappa$, and its real part
never exceeds the single-shell size.

\subsection{Heavy probes: a negative-temperature ensemble and the Nariai point instead of a singularity}
\label{sec:nariai}
Here we consider the case $\lambda\Delta=O(1)$, i.e.\ shells whose energy
$\omega\simeq\Delta/2$ is a finite fraction of $\sqrt{\kappa_0}$, the
scale at which the two-shell expansion
of~\cite{harlow2026antiscramblingeuclideanfoldsobserver} stops
applying.

DSSYK simplifies further when $\lambda\Delta$ is large. An $H$-chord crossing a matter
chord of dimension $\Delta$ costs $q^{\Delta}$; for $q^{\Delta}=e^{-\lambda\Delta}\to0$ the $H$-chords of each face must open and close
within the face, the vertex factors between adjacent faces go to one, and the four-point function factorises into a
product over faces, each a canonical ensemble at inverse
temperature equal to its gap. 

With $Z(g)=\int d\mu\,e^{gE}$ and given the $E\to-E$ symmetry of the measure,
each of the four faces contributes $Z(\betaf/2+it)$, so
$F_{\rm fold}\to q^{\Delta_V\Delta_W}\,Z(\betaf/2+it)^4$, where the
$t$-independent prefactor is the crossing of the two matter chords and
drops out of the face energies. This is a complex value whose phase winds
with $t$. In the same limit the heavy two-point function is
$G_0(\tau)\to Z(\tau)^2$, so $Z(\betaf/2+it)^4$ is the product of the
$VV$ and $WW$ two-point functions at separation $\betaf/2+it$. There is
no small connected correction to expand in:
the standard OTOC goes to $q^{\Delta^2}\to0$ at $t=0$, and the eikonal
regime $a\ll1$ is not available.

The folded face energy is
\begin{equation}
\langle E\rangle_{\rm fold}\;\xrightarrow{\;\lambda\Delta\gg1\;}\;
\frac{\int d\mu\,E\,e^{(\betaf/2+it)E}}{\int d\mu\,e^{(\betaf/2+it)E}},
\label{eq:canon}
\end{equation}
the thermal energy of DSSYK at inverse temperature $-\betaf/2$,
continued in real time. For $t\neq0$ the weights are complex, so this is
the logarithmic derivative of the partition function at complex $\beta$
rather than an ensemble average, which is why its real part is not
confined to $|E|\le E_0$. A negative-temperature form of the fold was
already noted in~\cite{harlow2026antiscramblingeuclideanfoldsobserver}
(their eqs.~(6.11), (6.12)); in DSSYK the $H\to-H$ symmetry of the
spectrum turns their (6.11) into the evenness of $f$ used in
Section~\ref{sec:fold}.
In the bulk the heavy
probes are walls that cut the geometry into static patches with no
interaction between them.

At $t=0$ the Boltzmann factor $e^{+\beta_fE/2}$ is already pushing
towards the upper spectral edge, where, however, the density of states
is vanishing like $\sin^2\theta$. 
The balance is determined by the saddle of
$\mathcal S(\theta)-\pi\cos\theta$, at $\theta=\pi/2+u$ with
$4u=\pi\cos u$, giving $E/E_0=\sin u=0.592$ as
$\lambda\to0$; the exact numerical measure gives $0.585$ at $q$$=$$0.95$. The folded
face thus sits more than half-way to $\kappa=0$ before the heavy shells
have moved.

\begin{figure}[htb]
\begin{center}
\includegraphics[scale=1.1]{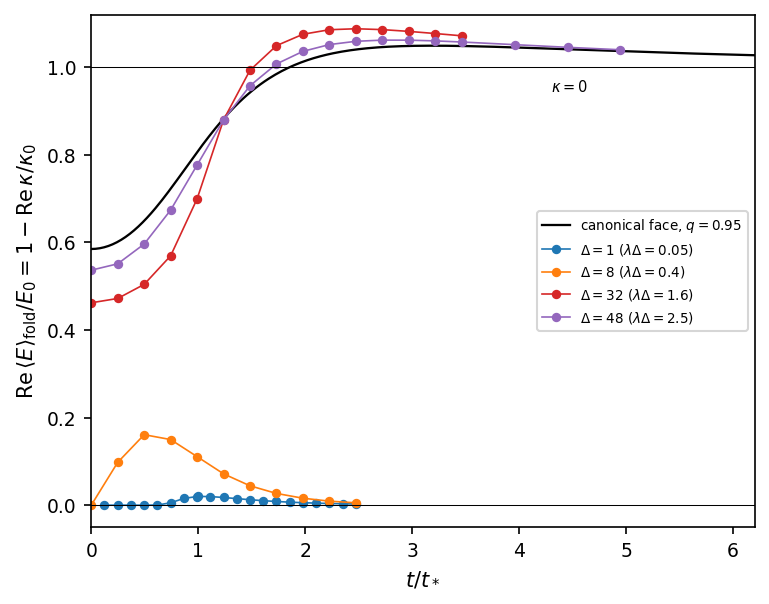}
\caption{
The DSSYK folded face energy at $q=0.95$ ($\lambda=0.051$) for light and heavy probes.
Plotted are the exact DSSYK values for ${\rm Re}\langle E\rangle_{\rm fold}/E_0=1-{\rm Re}\,\kappa/\kappa_0$ for $\Delta=1,8,32,48,$
($\lambda\Delta$=0.05, 0.4, 1.6, 2.5), against the canonical face (\ref{eq:canon}), which reaches the edge, Re$\,\kappa=0$, at
$t_c\simeq1.1\beta_f$ ($1.86t_*$ at this coupling), overshoots it by at most $5\%$
and returns to it as $\pi^2/(2\lambda t^2)$ for $t\gtrsim 5\,\betaf$.
Shells with $\lambda\Delta <1$ ($\Delta=1,8$) keep the light-shell form, a bounded peak followed by
relaxation; for $\lambda\Delta=1.6$ and $2.5$ the folded face starts at $(1-q^\Delta)$ times the canonical value and
follows the canonical curve to the edge.
The numerical results shown here required multiple precision
with $n_{\rm max}$ as high as 480 and dps as high as 360.
}
\label{fig:lateheavy}
\end{center}
\end{figure}

Figure~\ref{fig:lateheavy} illustrates the behavior of the
canonical face for $q=0.95$, along with exact DSSYK results for
$\Delta = $1,8,32, and 48, corresponding to $\lambda\Delta$ =
0.05, 0.4, 1.6, and 2.5. The canonical curve
crosses Re$\,\kappa=0$ at $t = 1.09\, \beta_f$, goes slightly negative,
then returns to zero from below as Re$\,\kappa/\kappa_0\simeq-\pi^2/(2\lambda t^2)$,
for $\betaf\ll t\ll\lambda^{-3/2}$.
The reason it can go slightly negative in spite of only
having support for $E \leq E_0$ is because of the
complex weights.

The two light examples shown in the figure $\Delta = $1 and 8 that have
$\lambda\Delta < 1$ both exhibit the light probe behavior
and do not cross zero. The two heavy examples, $\Delta = $32 and 48,
track the canonical curve up to corrections of order $O(q^\Delta)$.

The folded face and the two heavy examples thus reach the upper edge of the spectrum,
$\theta$$=$$\pi$, $E=E_0$, which is the Nariai point of the sine-dilaton
solutions (\ref{eq:sdsol}), where the black-hole and cosmological
horizons coincide. They reach it as the band edge of a negative-temperature
ensemble, at a time set by $\betaf$ and unrelated to the
scrambling time, and with no collision of probes. This is the sense in
which the bounded spectrum replaces the classical crunch: the quantity
that would diverge is the energy of a thermal ensemble, but it
saturates.


\section{Discussion and Outlook}\label{sec:discussion}

We have shown that the Euclidean-fold prescription of Harlow and Zhao,
implemented in the double-scaled SYK model at infinite temperature
with the fake temperature $\betaf=\pi/\Jf$ defining the fold, produces
a correlator that anti-scrambles: the connected part grows rather than decays.
We have shown that the exact DSSYK evaluation approximately tracks the
eikonal amplitude $\Phi(-a)$, with the shockwave time delay replaced by a time advance, through the scrambling time $t_*$ and beyond, out to $2.5\,t_*$ at $q=0.95$.

Because DSSYK is exactly solvable, the sheet selection can be checked
with no free parameter: the continued eikonal amplitude is complex, the
$n=0$ fold approaches the cut from below, and the exact correlator
carries the exponentially small term $\pi w^2(w-1)e^{-w}$ with the
positive sign that this predicts, growing into an $O(1)$ imaginary part
at the scrambling time; the $n=1$ fold gives the conjugate.

Using the face energies of the chord expansion we have compared the
folded correlator with the two-shell dilaton invariant
of~\cite{harlow2026antiscramblingeuclideanfoldsobserver}. The growing
term appears with the same exponent, but as an imaginary part of the
invariant, driven by the growth of the modulus of the folded
correlator. Its real part, which carries the backreaction in the
classical analysis, only moves by $O(\lambda)$ for thermal-scale probes,
the size of the single-shell terms, so
the fold does not reproduce the approach to the crunch. For heavy
probes the correlator factorizes and the folded face is a
negative-temperature ensemble whose energy saturates at the band edge,
the Nariai point of the sine-dilaton dual.

At finite $\beta$ the natural generalization is to fold by half the fake
period, $\beta/(2v)$. Modulo the effect of thermal
spreading of the fold rotation,
a folding prescription based on this idea gives reasonable agreement with the de Sitter eikonal form
for sufficiently high temperatures such that we satisfy the condition
that the backward Euclidean segment
is less than $b_c\simeq1.5$--$1.6\,\betaf/2$; at lower temperatures where this bound is
exceeded, the correlator is dominated by a
pair of complex conjugate saddles, an anti-Stokes crossover that prevents retrieval of
the eikonal behavior from the folded OTOC.

Our results are consistent with Chen, Stanford, Tang and Yang~\cite{chen2026negativeshocksversusstatic},
who show that the static-patch eikonal OTOC with observer recoil is not a
trace of Hermitian operators on a Hilbert space: for such a trace the
regularized OTOC cannot exceed its factorized value, and the static-patch
correlator does.
DSSYK at $\beta=0$ is such a trace. Its standard OTOC obeys
$|F_{\rm std}|\le1$ and decreases from $q^{\Delta_V\Delta_W}$ with exponent
$2\pi/\beta_f$.
The growth we find is in
$F_{\rm fold}(t)=\mathrm{Tr}[\tilde V(-t)W\tilde V(-t)W]$,
$\tilde V=e^{-\beta_fH/2}Ve^{\beta_fH/2}$, $\tilde V(t)=e^{iHt}\tilde Ve^{-iHt}$, a cyclic and positive trace of a
non-Hermitian probe whose Cauchy--Schwarz bound is
$\mathrm{Tr}[\tilde V^\dagger\tilde V]=G_0(\beta_f)$, exponentially large
in $1/\lambda$ ($G_0(\betaf)\simeq e^{3.2/\lambda}$ for $q=0.9$--$0.95$); the $O(1)$ value of $F_{\rm fold}$ is a cancellation fixed by
the fold being half the fake period.
The map from the de Sitter field to DSSYK is $V\mapsto V$ at two points and
$V\mapsto\tilde V$ at four points. It preserves the two-point KMS structure
and breaks it at four points, as 
in~\cite{harlow2026antiscramblingeuclideanfoldsobserver,cui2026sitterantiscramblingalgebra}, 
and it is not a $*$-map;
thus it does not contradict the tracial no-go.

Several questions are left open:
\begin{itemize}
\item Here we have for the most part taken the matter operator dimensions as $\Delta_V=\Delta_W=1$; as we have noted,
for general dimensions the
fake period itself depends on $\Delta$ at finite $\lambda$;
it would be good to understand this better conceptually.
\item The folded correlator is not the expectation value of a
product of operators in a state, so it is natural to ask if it
can be given an operational meaning, for example in a two-sided
version of the construction. A two-sided approach could also
connect to
the de Sitter OTOC analyses
of~\cite{Aalsma_2020,
milekhin2026outoftimeorderedcorrelatorssitterrevisited}.
\item It is natural to inquire if this entire analysis could
be uplifted to three dimensions; this would require making
a direct correspondence with the Virasoro TQFT
of~\cite{Collier_2023}, as
in \cite{chandra20263dsittercomplexvirasoro}.

\end{itemize}

\section*{Acknowledgements}
We thank Alex Zlokapa for useful discussions.
Madeline Mooney is supported by the Research Corporation for Science Advancement
and by the Brinson Foundation.
This work has been authored by Fermi Forward Discovery Group, LLC under Contract No.
89243024CSC000002 with the U.S. Department of Energy, Office of Science, Office of High Energy Physics.
This research was partly supported by the Quantum Communication Channels for Fundamental Physics (QCCFP)
QuantISED program grant, U.S. Department of Energy, Office of Science, Office of High Energy Physics, DE-SC0019219.


\bibliography{scibib}

@book{bender1999advanced,
  title={Advanced Mathematical Methods for Scientists and Engineers I: Asymptotic Methods and Perturbation Theory},
  author={Bender, Carl M. and Orszag, Steven A.},
  year={1999},
  publisher={Springer},
  address={New York, NY},
  doi={10.1007/978-1-4757-3069-2}
}

@article{Collier_2023,
   title={Solving 3d gravity with Virasoro TQFT},
   volume={15},
   ISSN={2542-4653},
   url={http://dx.doi.org/10.21468/SciPostPhys.15.4.151},
   DOI={10.21468/scipostphys.15.4.151},
   number={4},
   journal={SciPost Phys.},
   publisher={Stichting SciPost},
   author={Collier, Scott and Eberhardt, Lorenz and Zhang, Mengyang},
   year={2023},
   month=Oct, pages={151} }

@misc{chandra20263dsittercomplexvirasoro,
      title={{3D} de {Sitter} from complex {Virasoro} {TQFT}: Observers and knots}, 
      author={Jeevan Chandra and Diego Liska and Boris Post},
      year={2026},
      eprint={2610.30247},
      archivePrefix={arXiv},
      primaryClass={hep-th},
      url={https://arxiv.org/abs/2610.30247}, 
}

@article{okuyama2025sitterjtgravitydoublescaled,
      title={de {Sitter} {JT} gravity from double-scaled {SYK}},
      author={Kazumi Okuyama},
      journal={JHEP},
      volume={08},
      pages={181},
      year={2025},
      doi={10.1007/JHEP08(2025)181},
      eprint={2505.08116},
      archivePrefix={arXiv},
      primaryClass={hep-th},
}

@article{Dray:1984ha,
    author = "Dray, Tevian and 't Hooft, Gerard",
    title = "{The Gravitational Shock Wave of a Massless Particle}",
    reportNumber = "Print-84-0773 (UTRECHT)",
    doi = "10.1016/0550-3213(85)90525-5",
    journal = "Nucl. Phys. B",
    volume = "253",
    pages = "173--188",
    year = "1985"
}

@misc{harlow2026antiscramblingeuclideanfoldsobserver,
      title={Anti-scrambling and euclidean folds from observer correlators in de {Sitter} space}, 
      author={Daniel Harlow and Ying Zhao},
      year={2026},
      eprint={2607.14215},
      archivePrefix={arXiv},
      primaryClass={hep-th},
      url={https://arxiv.org/abs/2607.14215}, 
}

@misc{cui2026sitterantiscramblingalgebra,
      title={A de {Sitter} Anti-Scrambling Algebra}, 
      author={Wentao Cui and David K. Kolchmeyer},
      year={2026},
      eprint={2607.13665},
      archivePrefix={arXiv},
      primaryClass={hep-th},
      url={https://arxiv.org/abs/2607.13665}, 
}

@article{susskind2021entanglementchaossitterholography,
      title={Entanglement and Chaos in de {Sitter} Holography: An {SYK} Example},
      author={Leonard Susskind},
      journal={J. Hologr. Appl. Phys.},
      volume={1},
      number={1},
      pages={1--22},
      year={2021},
      doi={10.22128/jhap.2021.455.1005},
      eprint={2109.14104},
      archivePrefix={arXiv},
      primaryClass={hep-th},
}

@misc{susskind2022scramblingdoublescaledsyksitter,
      title={Scrambling in Double-Scaled {SYK} and de {Sitter} Space}, 
      author={Leonard Susskind},
      year={2022},
      eprint={2205.00315},
      archivePrefix={arXiv},
      primaryClass={hep-th},
      url={https://arxiv.org/abs/2205.00315}, 
}

@article{susskind2022sitterspacedoublescaledsyk,
      title={De {Sitter} Space, Double-Scaled {SYK}, and the Separation of Scales in the Semiclassical Limit},
      author={Leonard Susskind},
      journal={J. Hologr. Appl. Phys.},
      volume={5},
      number={1},
      pages={1--30},
      year={2025},
      doi={10.22128/jhap.2024.920.1103},
      eprint={2209.09999},
      archivePrefix={arXiv},
      primaryClass={hep-th},
}

@misc{rahman2024dsjtgravitydoublescaled,
      title={{dS} {JT} Gravity and Double-Scaled {SYK}}, 
      author={Adel A. Rahman},
      year={2022},
      eprint={2209.09997},
      archivePrefix={arXiv},
      primaryClass={hep-th},
      url={https://arxiv.org/abs/2209.09997}, 
}

@misc{rahman2025infinitetemperatureinfinitetemperatures,
      title={Infinite Temperature is Not So Infinite: The Many Temperatures of de {Sitter} Space}, 
      author={Adel A. Rahman and Leonard Susskind},
      year={2024},
      eprint={2401.08555},
      archivePrefix={arXiv},
      primaryClass={hep-th},
      url={https://arxiv.org/abs/2401.08555}, 
}

@article{narovlansky2023doublescaledsyksitterholography,
      title={Double-scaled {SYK} and de {Sitter} Holography},
      author={Vladimir Narovlansky and Herman Verlinde},
      journal={JHEP},
      volume={05},
      pages={032},
      year={2025},
      doi={10.1007/JHEP05(2025)032},
      eprint={2310.16994},
      archivePrefix={arXiv},
      primaryClass={hep-th},
}

@article{Blommaert_2025,
   title={The dilaton gravity hologram of double-scaled SYK},
   volume={06},
   ISSN={1029-8479},
   url={http://dx.doi.org/10.1007/JHEP06(2025)050},
   DOI={10.1007/jhep06(2025)050},
   journal={JHEP},
   publisher={Springer Science and Business Media LLC},
   author={Blommaert, Andreas and Mertens, Thomas G. and Papalini, Jacopo},
   year={2025},
   month=Jun, pages={050} }

@article{Lin_2023,
   title={A symmetry algebra in double-scaled SYK},
   volume={15},
   ISSN={2542-4653},
   url={http://dx.doi.org/10.21468/SciPostPhys.15.6.234},
   DOI={10.21468/scipostphys.15.6.234},
   number={6},
   journal={SciPost Phys.},
   publisher={Stichting SciPost},
   author={Lin, Henry W. and Stanford, Douglas},
   year={2023},
   month=Dec, pages={234} }

@misc{lin2022infinitetemperatureshot,
      title={Infinite Temperature's Not So Hot}, 
      author={Henry Lin and Leonard Susskind},
      year={2022},
      eprint={2206.01083},
      archivePrefix={arXiv},
      primaryClass={hep-th},
      url={https://arxiv.org/abs/2206.01083}, 
}

@article{Aalsma_2020,
   title={Chaos and complementarity in de Sitter space},
   volume={05},
   ISSN={1029-8479},
   url={http://dx.doi.org/10.1007/JHEP05(2020)152},
   DOI={10.1007/jhep05(2020)152},
   journal={JHEP},
   publisher={Springer Science and Business Media LLC},
   author={Aalsma, Lars and Shiu, Gary},
   year={2020},
   month=May, pages={152} }

@misc{milekhin2026outoftimeorderedcorrelatorssitterrevisited,
      title={Out-of-time-ordered Correlators in de {Sitter} Revisited}, 
      author={Alexey Milekhin and Vladimir Narovlansky and Jiuci Xu},
      year={2026},
      eprint={2607.13137},
      archivePrefix={arXiv},
      primaryClass={hep-th},
      url={https://arxiv.org/abs/2607.13137}, 
}

@misc{chen2026negativeshocksversusstatic,
      title={Negative shocks versus static patch holography}, 
      author={Yiming Chen and Douglas Stanford and Haifeng Tang and Zhenbin Yang},
      year={2026},
      eprint={2607.14042},
      archivePrefix={arXiv},
      primaryClass={hep-th},
      url={https://arxiv.org/abs/2607.14042}, 
}

@misc{witten2001quantumgravitysitterspace,
      title={Quantum Gravity In de {Sitter} Space}, 
      author={Edward Witten},
      year={2001},
      eprint={hep-th/0106109},
      archivePrefix={arXiv},
      url={https://arxiv.org/abs/hep-th/0106109}, 
}

@article{Strominger_2001,
   title={The dS/CFT correspondence},
   volume={10},
   ISSN={1029-8479},
   url={http://dx.doi.org/10.1088/1126-6708/2001/10/034},
   DOI={10.1088/1126-6708/2001/10/034},
   journal={JHEP},
   publisher={Springer Science and Business Media LLC},
   author={Strominger, Andrew},
   year={2001},
   month=Oct, pages={034} }

@article{Maldacena_2003,
   title={Non-gaussian features of primordial fluctuations in single field inflationary models},
   volume={05},
   ISSN={1029-8479},
   url={http://dx.doi.org/10.1088/1126-6708/2003/05/013},
   DOI={10.1088/1126-6708/2003/05/013},
   journal={JHEP},
   publisher={Springer Science and Business Media LLC},
   author={Maldacena, Juan},
   year={2003},
   month=May, pages={013} }

@article{xu2026vonneumannalgebrasdoublescaled,
      title={Von {Neumann} Algebras in Double-Scaled {SYK}},
      author={Jiuci Xu},
      journal={JHEP},
      volume={03},
      pages={016},
      year={2026},
      doi={10.1007/JHEP03(2026)016},
      eprint={2403.09021},
      archivePrefix={arXiv},
      primaryClass={hep-th},
}

@article{garcia2016spectral,
  title={Spectral and thermodynamic properties of the Sachdev-Ye-Kitaev model},
  author={Garc\'{\i}a-Garc\'{\i}a, Antonio M. and Verbaarschot, Jacobus J. M.},
  journal={Phys. Rev. D},
  volume={94},
  number={12},
  pages={126010},
  year={2016},
  publisher={APS},
  doi={10.1103/PhysRevD.94.126010}
}

@Article{Cotler2017,
author={Cotler, Jordan S.
and Gur-Ari, Guy
and Hanada, Masanori
and Polchinski, Joseph
and Saad, Phil
and Shenker, Stephen H.
and Stanford, Douglas
and Streicher, Alexandre
and Tezuka, Masaki},
title={Black holes and random matrices},
journal={JHEP},
year={2017},
month={May},
day={22},
volume={05},
pages={118},
issn={1029-8479},
doi={10.1007/JHEP05(2017)118},
url={https://doi.org/10.1007/JHEP05(2017)118}
}

@article{Sachdev:2015efa,
    author = "Sachdev, Subir",
    title = "{Bekenstein-Hawking Entropy and Strange Metals}",
    eprint = "1506.05111",
    archivePrefix = "arXiv",
    primaryClass = "hep-th",
    doi = "10.1103/PhysRevX.5.041025",
    journal = "Phys. Rev. X",
    volume = "5",
    number = "4",
    pages = "041025",
    year = "2015"
}

@article{Berkooz:2020uly,
    author = "Berkooz, Micha and Narovlansky, Vladimir and Raj, Himanshu",
    title = "{Complex Sachdev-Ye-Kitaev model in the double scaling limit}",
    eprint = "2006.13983",
    archivePrefix = "arXiv",
    primaryClass = "hep-th",
    doi = "10.1007/JHEP02(2021)113",
    journal = "JHEP",
    volume = "02",
    pages = "113",
    year = "2021"
}

@article{Maldacena:2016hyu,
    author = "Maldacena, Juan and Stanford, Douglas",
    title = "{Remarks on the Sachdev-Ye-Kitaev model}",
    eprint = "1604.07818",
    archivePrefix = "arXiv",
    primaryClass = "hep-th",
    doi = "10.1103/PhysRevD.94.106002",
    journal = "Phys. Rev. D",
    volume = "94",
    number = "10",
    pages = "106002",
    year = "2016"
}

@article{Maldacena:2001kr,
    author = "Maldacena, Juan Martin",
    title = "{Eternal black holes in anti-de Sitter}",
    eprint = "hep-th/0106112",
    archivePrefix = "arXiv",
    reportNumber = "NSF-ITP-01-59",
    doi = "10.1088/1126-6708/2003/04/021",
    journal = "JHEP",
    volume = "04",
    pages = "021",
    year = "2003"
}

@article{Son_2002,
   title={Minkowski-space correlators in AdS/CFT correspondence: recipe and applications},
   volume={09},
   ISSN={1029-8479},
   url={http://dx.doi.org/10.1088/1126-6708/2002/09/042},
   DOI={10.1088/1126-6708/2002/09/042},
   journal={JHEP},
   publisher={Springer Science and Business Media LLC},
   author={Son, Dam T and Starinets, Andrei O},
   year={2002},
   month=Sep, pages={042} }

@article{Roberts_2015,
   title={Diagnosing Chaos Using Four-Point Functions in Two-Dimensional Conformal Field Theory},
   volume={115},
   ISSN={1079-7114},
   url={http://dx.doi.org/10.1103/PhysRevLett.115.131603},
   DOI={10.1103/physrevlett.115.131603},
   number={13},
   journal={Phys. Rev. Lett.},
   publisher={American Physical Society (APS)},
   author={Roberts, Daniel A. and Stanford, Douglas},
   year={2015},
   month=Sep, pages={131603} }

@article{Shenker:2013pqa,
    author = "Shenker, Stephen H. and Stanford, Douglas",
    title = "{Black holes and the butterfly effect}",
    eprint = "1306.0622",
    archivePrefix = "arXiv",
    primaryClass = "hep-th",
    reportNumber = "SU-ITP-13-08",
    doi = "10.1007/JHEP03(2014)067",
    journal = "JHEP",
    volume = "03",
    pages = "067",
    year = "2014"
}

@article{shenker2015stringyeffectsscrambling,
      title={Stringy effects in scrambling},
      author={Stephen H. Shenker and Douglas Stanford},
      journal={JHEP},
      volume={05},
      pages={132},
      year={2015},
      doi={10.1007/JHEP05(2015)132},
      eprint={1412.6087},
      archivePrefix={arXiv},
      primaryClass={hep-th},
}

@article{gao2017traversable,
  title={Traversable wormholes via a double trace deformation},
  author={Gao, Ping and Jafferis, Daniel Louis and Wall, Aron C},
  journal={JHEP},
  volume={12},
  pages={151},
  year={2017},
  publisher={Springer},
  doi={10.1007/JHEP12(2017)151}
}

@article{PhysRevLett.70.3339,
  title = {Gapless spin-fluid ground state in a random quantum Heisenberg magnet},
  author = {Sachdev, Subir and Ye, Jinwu},
  journal = {Phys. Rev. Lett.},
  volume = {70},
  issue = {21},
  pages = {3339--3342},
  numpages = {0},
  year = {1993},
  month = {May},
  publisher = {American Physical Society},
  doi = {10.1103/PhysRevLett.70.3339},
  url = {https://link.aps.org/doi/10.1103/PhysRevLett.70.3339}
}

@misc{kitaev2015simple,
  title={A simple model of quantum holography},
  author={Kitaev, Alexei},
  howpublished={Talks at KITP, April 7 and May 27},
  year={2015},
  note={\url{https://online.kitp.ucsb.edu/online/entangled15/}}
}

@Article{maldacena2016bound,
author={Maldacena, Juan
and Shenker, Stephen H.
and Stanford, Douglas},
title={A bound on chaos},
journal={JHEP},
year={2016},
volume={08},
pages={106},
issn={1029-8479},
doi={10.1007/JHEP08(2016)106},
url={https://doi.org/10.1007/JHEP08(2016)106}
}

@article{Berkooz_2018,
   title={Chord diagrams, exact correlators in spin glasses and black hole bulk reconstruction},
   volume={08},
   ISSN={1029-8479},
   url={http://dx.doi.org/10.1007/JHEP08(2018)192},
   DOI={10.1007/jhep08(2018)192},
   journal={JHEP},
   publisher={Springer Science and Business Media LLC},
   author={Berkooz, Micha and Narayan, Prithvi and Sim\'{o}n, Joan},
   year={2018},
   month=Aug, pages={192} }

@article{Berkooz:2018jqr,
    author = "Berkooz, Micha and Isachenkov, Mikhail and Narovlansky, Vladimir and Torrents, Genis",
    title = "{Towards a full solution of the large N double-scaled SYK model}",
    eprint = "1811.02584",
    archivePrefix = "arXiv",
    primaryClass = "hep-th",
    doi = "10.1007/JHEP03(2019)079",
    journal = "JHEP",
    volume = "03",
    pages = "079",
    year = "2019"
}

@article{berkooz2024cordialintroductiondoublescaled,
      title={A Cordial Introduction to Double Scaled {SYK}},
      author={Micha Berkooz and Ohad Mamroud},
      journal={Rept. Prog. Phys.},
      volume={88},
      number={3},
      pages={036001},
      year={2025},
      doi={10.1088/1361-6633/ada889},
      eprint={2407.09396},
      archivePrefix={arXiv},
      primaryClass={hep-th},
}

@article{mukhametzhanov2023largepsykchord,
      title={Large $p$ {SYK} from chord diagrams},
      author={Baur Mukhametzhanov},
      journal={JHEP},
      volume={09},
      pages={154},
      year={2023},
      doi={10.1007/JHEP09(2023)154},
      eprint={2303.03474},
      archivePrefix={arXiv},
      primaryClass={hep-th},
}

@Article{Kitaev2018,
author={Kitaev, Alexei
and Suh, S. Josephine},
title={The soft mode in the Sachdev-Ye-Kitaev model and its gravity dual},
journal={JHEP},
year={2018},
month={May},
day={29},
volume={05},
pages={183},
issn={1029-8479},
doi={10.1007/JHEP05(2018)183},
url={https://doi.org/10.1007/JHEP05(2018)183}
}

@article{Roberts:2018mnp,
    author = "Roberts, Daniel A. and Stanford, Douglas and Streicher, Alexandre",
    title = "{Operator growth in the SYK model}",
    eprint = "1802.02633",
    archivePrefix = "arXiv",
    primaryClass = "hep-th",
    doi = "10.1007/JHEP06(2018)122",
    journal = "JHEP",
    volume = "06",
    pages = "122",
    year = "2018"
}

@article{PhysRevD.96.066012,
  title = {Analytical spectral density of the Sachdev-Ye-Kitaev model at finite $N$},
  author = {Garc\'{\i}a-Garc\'{\i}a, Antonio M. and Verbaarschot, Jacobus J. M.},
  journal = {Phys. Rev. D},
  volume = {96},
  issue = {6},
  pages = {066012},
  numpages = {10},
  year = {2017},
  month = {Sep},
  publisher = {American Physical Society},
  doi = {10.1103/PhysRevD.96.066012},
  url = {https://link.aps.org/doi/10.1103/PhysRevD.96.066012}
}

@article{Maldacena:2016upp,
    author = "Maldacena, Juan and Stanford, Douglas and Yang, Zhenbin",
    title = "{Conformal symmetry and its breaking in two dimensional Nearly Anti-de-Sitter space}",
    eprint = "1606.01857",
    archivePrefix = "arXiv",
    primaryClass = "hep-th",
    doi = "10.1093/ptep/ptw124",
    journal = "PTEP",
    volume = "2016",
    number = "12",
    pages = "12C104",
    year = "2016"
}

%
%
%
\makeatletter
\@ifundefined{ruledtabular}{%
  \newenvironment{ruledtabular}{}{}%
  \@ifundefined{toprule}{%
    \newcommand{\rtTop}{\hline\hline}\newcommand{\rtBot}{\hline\hline}%
    \providecommand{\colrule}{\hline}%
  }{%
    \newcommand{\rtTop}{\toprule}\newcommand{\rtBot}{\bottomrule}%
    \providecommand{\colrule}{\midrule}%
  }%
}{%
  \newcommand{\rtTop}{}\newcommand{\rtBot}{}%
}
\makeatother

\begin{table}[htbp]
  \centering
  \renewcommand{\arraystretch}{0.90} 
  \caption{Fold-amplitude data for $q=0.975$
    ($\lambda=0.0253$, $2\sqrt{\lambda}=0.318$, $\beta_f=19.683$, $t_*=13.74$).
    $\delta=|\Phi(-a)-F_{\mathrm{fold}}|/|F_{\mathrm{fold}}-1|$ is the relative deviation of the complex amplitudes.}
  \label{tab:q975}
  \begin{ruledtabular}
  \begin{tabular}{
      S[table-format=2.2]
      S[table-format=1.5]
      S[table-format=1.3]
      S[table-format=+1.4]
      S[table-format=+1.4]
      S[table-format=1.4]
      S[table-format=+1.3]
      S[table-format=+1.3]
    }
    \rtTop
    {$t$} & {$a(t)$} & {$\mathrm{d}\ln a/\mathrm{d}t$}
      & {$\operatorname{Re}F_{\mathrm{fold}}-1$} & {$\operatorname{Re}\Phi(-a)-1$}
      & {$\delta$} & {$\operatorname{Im}F_{\mathrm{fold}}$} & {$\operatorname{Im}\Phi(-a)$} \\
    \colrule
    0.00 & 0.00643 & {---} & +0.0256 & +0.0265 & 0.0329 & 0.000 & 0.000 \\
    0.50 & 0.00651 & 0.025 & +0.0260 & +0.0268 & 0.0325 & 0.000 & 0.000 \\
    1.00 & 0.00675 & 0.074 & +0.0270 & +0.0279 & 0.0313 & 0.000 & 0.000 \\
    1.50 & 0.00717 & 0.119 & +0.0288 & +0.0296 & 0.0295 & 0.000 & 0.000 \\
    2.00 & 0.00776 & 0.159 & +0.0313 & +0.0322 & 0.0272 & 0.000 & 0.000 \\
    2.50 & 0.00855 & 0.194 & +0.0347 & +0.0356 & 0.0246 & 0.000 & 0.000 \\
    3.00 & 0.00956 & 0.222 & +0.0391 & +0.0400 & 0.0220 & 0.000 & 0.000 \\
    3.50 & 0.01081 & 0.245 & +0.0446 & +0.0455 & 0.0194 & 0.000 & 0.000 \\
    4.00 & 0.01233 & 0.263 & +0.0514 & +0.0522 & 0.0169 & 0.000 & 0.000 \\
    4.50 & 0.01416 & 0.277 & +0.0597 & +0.0606 & 0.0146 & 0.000 & 0.000 \\
    5.00 & 0.01635 & 0.288 & +0.0698 & +0.0707 & 0.0125 & 0.000 & 0.000 \\
    5.50 & 0.01896 & 0.296 & +0.0822 & +0.0830 & 0.0107 & 0.000 & 0.000 \\
    6.00 & 0.02204 & 0.302 & +0.0973 & +0.0981 & 0.0089 & 0.000 & 0.000 \\
    6.50 & 0.02569 & 0.306 & +0.1157 & +0.1166 & 0.0074 & 0.000 & 0.000 \\
    7.00 & 0.02999 & 0.309 & +0.1385 & +0.1394 & 0.0059 & 0.000 & 0.000 \\
    7.50 & 0.03505 & 0.312 & +0.1669 & +0.1677 & 0.0044 & 0.000 & 0.000 \\
    8.00 & 0.04100 & 0.314 & +0.2028 & +0.2034 & 0.0029 & 0.000 & 0.000 \\
    8.50 & 0.04799 & 0.315 & +0.2489 & +0.2492 & 0.0012 & 0.000 & 0.000 \\
    9.00 & 0.05621 & 0.316 & +0.3102 & +0.3099 & 0.0012 & 0.000 & 0.000 \\
    9.50 & 0.06585 & 0.317 & +0.3954 & +0.3938 & 0.0045 & +0.003 & +0.003 \\
    10.00 & 0.07717 & 0.317 & +0.5182 & +0.5152 & 0.0084 & +0.018 & +0.015 \\
    10.50 & 0.09045 & 0.318 & +0.6935 & +0.6908 & 0.0122 & +0.069 & +0.061 \\
    11.00 & 0.10603 & 0.318 & +0.9214 & +0.9234 & 0.0149 & +0.203 & +0.189 \\
    11.50 & 0.12431 & 0.318 & +1.1630 & +1.1753 & 0.0162 & +0.476 & +0.460 \\
    12.00 & 0.14575 & 0.318 & +1.3314 & +1.3559 & 0.0165 & +0.919 & +0.908 \\
    12.50 & 0.17090 & 0.318 & +1.3168 & +1.3488 & 0.0161 & +1.498 & +1.501 \\
    13.00 & 0.20039 & 0.318 & +1.0420 & +1.0715 & 0.0152 & +2.104 & +2.124 \\
    13.50 & 0.23498 & 0.319 & +0.5084 & +0.5258 & 0.0141 & +2.595 & +2.627 \\
    13.74 & 0.25365 & 0.319 & +0.1831 & +0.1928 & 0.0135 & +2.752 & +2.787 \\
    \rtBot
  \end{tabular}
  \end{ruledtabular}
\end{table}

\begin{table}[htbp]
  \centering
  \caption{Fold-amplitude data for $q=0.950$
    ($\lambda=0.0513$, $2\sqrt{\lambda}=0.453$, $\beta_f=13.786$, $t_*=8.09$).
    $\delta=|\Phi(-a)-F_{\mathrm{fold}}|/|F_{\mathrm{fold}}-1|$ is the relative deviation of the complex amplitudes.}
  \label{tab:q950}
  \begin{ruledtabular}
  \begin{tabular}{
      S[table-format=1.2]
      S[table-format=1.5]
      S[table-format=1.3]
      S[table-format=+1.4]
      S[table-format=+1.4]
      S[table-format=1.4]
      S[table-format=+1.3]
      S[table-format=+1.3]
    }
    \rtTop
    {$t$} & {$a(t)$} & {$\mathrm{d}\ln a/\mathrm{d}t$}
      & {$\operatorname{Re}F_{\mathrm{fold}}-1$} & {$\operatorname{Re}\Phi(-a)-1$}
      & {$\delta$} & {$\operatorname{Im}F_{\mathrm{fold}}$} & {$\operatorname{Im}\Phi(-a)$} \\
    \colrule
    0.00 & 0.01324 & {---} & +0.0527 & +0.0563 & 0.0697 & 0.000 & 0.000 \\
    0.50 & 0.01357 & 0.075 & +0.0542 & +0.0579 & 0.0679 & 0.000 & 0.000 \\
    1.00 & 0.01459 & 0.167 & +0.0588 & +0.0625 & 0.0630 & 0.000 & 0.000 \\
    1.50 & 0.01635 & 0.246 & +0.0669 & +0.0707 & 0.0560 & 0.000 & 0.000 \\
    2.00 & 0.01894 & 0.308 & +0.0791 & +0.0830 & 0.0481 & 0.000 & 0.000 \\
    2.50 & 0.02249 & 0.355 & +0.0965 & +0.1003 & 0.0401 & 0.000 & 0.000 \\
    3.00 & 0.02719 & 0.387 & +0.1205 & +0.1244 & 0.0327 & 0.000 & 0.000 \\
    3.50 & 0.03329 & 0.410 & +0.1537 & +0.1576 & 0.0259 & 0.000 & 0.000 \\
    4.00 & 0.04109 & 0.425 & +0.2001 & +0.2039 & 0.0194 & 0.000 & 0.000 \\
    4.50 & 0.05101 & 0.435 & +0.2672 & +0.2705 & 0.0124 & 0.000 & 0.000 \\
    5.00 & 0.06356 & 0.441 & +0.3710 & +0.3724 & 0.0045 & +0.003 & +0.002 \\
    5.50 & 0.07939 & 0.446 & +0.5431 & +0.5422 & 0.0123 & +0.026 & +0.020 \\
    6.00 & 0.09931 & 0.448 & +0.8167 & +0.8220 & 0.0240 & +0.142 & +0.122 \\
    6.50 & 0.12437 & 0.450 & +1.1465 & +1.1761 & 0.0308 & +0.485 & +0.461 \\
    7.00 & 0.15586 & 0.452 & +1.3252 & +1.3820 & 0.0324 & +1.146 & +1.145 \\
    7.50 & 0.19540 & 0.452 & +1.0798 & +1.1345 & 0.0309 & +1.986 & +2.029 \\
    8.09 & 0.25526 & 0.453 & +0.1548 & +0.1639 & 0.0272 & +2.724 & +2.798 \\
    \rtBot
  \end{tabular}
  \end{ruledtabular}
\end{table}

\begin{table}[htbp]
  \centering
  \caption{Fold-amplitude data for $q=0.900$
    ($\lambda=0.1054$, $2\sqrt{\lambda}=0.649$, $\beta_f=9.563$, $t_*=4.53$).
    $\delta=|\Phi(-a)-F_{\mathrm{fold}}|/|F_{\mathrm{fold}}-1|$ is the relative deviation of the complex amplitudes.}
  \label{tab:q900}
  \begin{ruledtabular}
  \begin{tabular}{
      S[table-format=1.2]
      S[table-format=1.5]
      S[table-format=1.3]
      S[table-format=+1.4]
      S[table-format=+1.4]
      S[table-format=1.4]
      S[table-format=+1.3]
      S[table-format=+1.3]
    }
    \rtTop
    {$t$} & {$a(t)$} & {$\mathrm{d}\ln a/\mathrm{d}t$}
      & {$\operatorname{Re}F_{\mathrm{fold}}-1$} & {$\operatorname{Re}\Phi(-a)-1$}
      & {$\delta$} & {$\operatorname{Im}F_{\mathrm{fold}}$} & {$\operatorname{Im}\Phi(-a)$} \\
    \colrule
    0.00 & 0.02810 & {---} & +0.1115 & +0.1292 & 0.1584 & 0.000 & 0.000 \\
    0.25 & 0.02845 & 0.050 & +0.1134 & +0.1311 & 0.1563 & 0.000 & 0.000 \\
    0.50 & 0.02953 & 0.148 & +0.1190 & +0.1369 & 0.1502 & 0.000 & 0.000 \\
    0.75 & 0.03135 & 0.240 & +0.1287 & +0.1468 & 0.1408 & 0.000 & 0.000 \\
    1.00 & 0.03398 & 0.321 & +0.1430 & +0.1615 & 0.1293 & 0.000 & 0.000 \\
    1.25 & 0.03747 & 0.391 & +0.1629 & +0.1818 & 0.1164 & 0.000 & 0.000 \\
    1.50 & 0.04192 & 0.449 & +0.1896 & +0.2091 & 0.1030 & 0.000 & 0.000 \\
    1.75 & 0.04745 & 0.496 & +0.2253 & +0.2454 & 0.0893 & 0.000 & 0.000 \\
    2.00 & 0.05421 & 0.533 & +0.2738 & +0.2943 & 0.0748 & 0.000 & 0.000 \\
    2.25 & 0.06237 & 0.561 & +0.3414 & +0.3617 & 0.0593 & +0.002 & +0.001 \\
    2.50 & 0.07217 & 0.583 & +0.4383 & +0.4582 & 0.0460 & +0.011 & +0.007 \\
    2.75 & 0.08385 & 0.600 & +0.5770 & +0.5995 & 0.0430 & +0.043 & +0.032 \\
    3.00 & 0.09773 & 0.613 & +0.7636 & +0.7981 & 0.0518 & +0.130 & +0.109 \\
    3.25 & 0.11418 & 0.622 & +0.9822 & +1.0427 & 0.0627 & +0.317 & +0.294 \\
    3.50 & 0.13364 & 0.629 & +1.1794 & +1.2733 & 0.0699 & +0.646 & +0.642 \\
    3.75 & 0.15663 & 0.635 & +1.2674 & +1.3824 & 0.0724 & +1.121 & +1.164 \\
    4.00 & 0.18376 & 0.639 & +1.1549 & +1.2581 & 0.0710 & +1.688 & +1.790 \\
    4.25 & 0.21576 & 0.642 & +0.7936 & +0.8480 & 0.0670 & +2.232 & +2.381 \\
    4.53 & 0.25843 & 0.645 & +0.1294 & +0.1069 & 0.0607 & +2.657 & +2.817 \\
    \rtBot
  \end{tabular}
  \end{ruledtabular}
\end{table}


\end{document}